# Preserving Metamagnetism in Self-Assembled FeRh Nanomagnets

Lucie Motyčková[1,2,†], Jon Ander Arregi[1,†,*], Michal Staňo[1], Stanislav Průša[1,2], Klára Částková[1,3], and Vojtěch Uhlíř[1,2,*]

[1]CEITEC BUT, Brno University of Technology, Purkyňova 123, 612 00 Brno, Czech Republic

[2]Institute of Physical Engineering, Brno University of Technology, Technická 2, 61669 Brno, Czech Republic

[3]Department of Ceramics and Polymers, Brno University of Technology, Technická 2, 616 69 Brno, Czech Republic

[†]These authors contributed equally to this work.

*Corresponding authors: ja.arregi@ceitec.vutbr.cz, vojtech.uhlir@ceitec.vutbr.cz

We present self-assembly routes of metamagnetic FeRh nanoisland arrays using sputter deposition processing. Controlling the growth parameters and epitaxy allows tuning the size and shape of the FeRh nanoislands. While the phase transition between antiferromagnetic and ferromagnetic order is largely suppressed in nanoislands formed on oxide substrates via thermodynamic nucleation, we find that metamagnetism is largely preserved in nanomagnet arrays formed through solid-state dewetting. This behavior is strongly dependent on the resulting crystal faceting of the nanoislands, which is characteristic of each assembly route. Comparing the calculated surface energies for each magnetic phase of the nanoislands reveals that metamagnetism can be suppressed or allowed by specific geometrical configurations of the facets. Furthermore, we find that the spatial confinement leads to very pronounced supercooling and the absence of phase separation in the nanoislands. Finally, the supported nanomagnets are chemically etched away from the substrates to inspect the phase transition properties of self-standing nanoparticles. We demonstrate that solid-state dewetting is a feasible and scalable way to obtain supported and free-standing FeRh nanomagnets with preserved metamagnetism.

# 1. Introduction

An ever-present challenge in nanotechnology is the large-scale synthesis of high-quality functional nanostructures with controlled size, shape, and properties (e.g., electronic, optical, magnetic, chemical) [1]. Self-assembly methods nowadays constitute a particularly efficient tool facilitating high-throughput fabrication of technologically relevant materials, such as catalysts [2], fuel cells [3], metal-organic frameworks [4], and optical metamaterials [5]. In the domain of magnetic materials, self-assembly also plays a crucial role in the fabrication of superparamagnetic iron-oxide nanoparticles [6], and granular recording media [7-9].

While nanostructuring can beneficially lead to emergent phenomena and novel functionalities, it is sometimes desirable that nanomaterials preserve specific bulk properties. In the case of phase change materials, nanoscale confinement and nanofabrication processing often lead to the unwanted suppression of certain electronic or magnetic ordering states, thus making phase transitions present in the bulk vanish or degrade. For instance, the metal-insulator transition in $VO_2$ is mitigated in ultrathin films and nanostructures, where the number of resistivity variation decades is reduced in comparison to the bulk [10-13]. The deterioration of bulk-like properties upon nanostructuring can be particularly severe in materials with interconnected structural, electronic, and magnetic order parameters. Factors such as excess strain, composition inhomogeneities, grain size effects, or lithographically induced defects can restrain intrinsic material functionalities [14, 15].

Here, we focus on the iron-rhodium (FeRh) alloy, a metallic system featuring a metamagnetic phase transition from antiferromagnetic (AF) to ferromagnetic (FM) order above room temperature ($T_M$ ~ 360 K) [16, 17]. The phase transition is first-order in nature, only exists in a narrow region near the equiatomic composition range for the CsCl-type structure, and presents a thermal hysteresis of around 10 K [18]. The sharp magnetization increase upon heating is accompanied by a concomitant isotropic lattice expansion (~0.5%) [19] and a reduction in resistivity (~50%) [20]. Over the last decades, FeRh has been studied as a test-bed for exploring the fundamental physics of coupled order parameters [21]. The large changes in magnetization, magnetoresistance and entropy, together with the option to control these changes via various driving forces (e.g., temperature, magnetic field, strain, light pulses), also make this material interesting for technological applications. FeRh has been proposed for incorporation into magnetic recording [22] or spintronic [23, 24] devices, and is utilized as a model platform for solid-state refrigeration technologies [25-29]. More recently, FeRh has been considered as a switchable high-contrast label for magnetic resonance imaging with the potential to work near the body temperature [30, 31].

Several fabrication routes have been explored for the self-assembly of nanoscale FeRh elements. On the one hand, solution-phase chemical methods lead to nanoparticles with sizes between 3-20 nm [32, 33]. Typically, only a minor fraction of the synthesized sample undergoes the phase transition, which is likely caused by the presence of fcc-ordered FeRh where the transition is inherently absent [32, 33]. More recently, Cao *et al.* reported the fabrication of bcc-like particulate FeRh alloys with a more prominent AF-FM phase transition, although the residual magnetization at low temperatures was still relatively large [34]. Additionally, Biswas *et al*. have succeeded in obtaining an abrupt phase transition in FeRh powders, which feature interconnected particles of 0.6-1 μm in size [35].

On the other hand, FeRh nanoislands with sizes in the range of 10-100 nm and supported on crystal substrates have been fabricated using self-organization during physical vapor deposition. This strategy exploits the Volmer-Weber growth mode during high-temperature deposition of FeRh on single-crystal MgO, resulting in the nucleation of physically separated epitaxial islands [36-38]. Despite the bcc-like structure being favorably imposed via epitaxy, it was concluded that the FM stabilized surface shell impedes the AF state at the nanoisland core, suppressing the phase transition [37, 38].

These findings underline the exceptionally large sensitivity of metamagnetism in FeRh to different factors, where the phase transition characteristics are affected by stoichiometry, strain and defects [19, 39], the existence of residual FM-stabilized regions at the interfaces [40-42], or nanoscale morphology [38]. These observations highlight the need for alternative self-assembly routes of phase change materials to preserve functionalities upon nanoscale size confinement.

In this work, we present the self-assembly of epitaxial sub-micron FeRh nanomagnets with preserved metamagnetism using solid-state dewetting [43-45]. Starting from thin epitaxial FeRh films on single-crystal substrates, we fabricate arrays of FeRh nanoislands with tunable sizes and shapes. The nanoislands assembled via dewetting sustain the AF-FM phase transition, in contrast to those of comparable size originating from the Volmer-Weber nucleation. We investigate the morphological features of the islands formed upon each assembly route, finding a strong correlation between the existence of metamagnetism and dominant crystal faceting of the nanoislands. This identifies morphology as the leading factor suppressing or allowing the phase transition in FeRh nanoislands. The magnetic phase transition in individual FeRh nanoislands presents extraordinary size-dependent effects, such as extended supercooling (>150 K). Finally, the nanoislands are chemically etched from the substrates to obtain metamagnetic FeRh nanoparticles in solution.

## 2. Results and Discussion

### 2.1 Self-assembly of FeRh nanoislands

We report two distinct processes that combine magnetron sputter deposition and annealing leading to self-assembly of sub-micron FeRh islands on single-crystal MgO(001), MgO(011) and Al$_2$O$_3$(0001) substrates. Both routes are triggered by thermodynamic surface energy minimization and driven by high-temperature annealing during and/or after the growth. For the sake of simplicity, we initially describe the MgO(001) substrate system.

In the first process, FeRh growth is initiated at 750 K with a deposition rate of 2 nm min$^{-1}$. After 3 min of deposition, the substrate temperature is ramped within 5 min to 1100 K, maintaining this temperature until completing the film with a nominal thickness $t$. The sample is subsequently annealed at 1100 K for 80 min. This process does not lead to the formation of a continuous FeRh film, but rather results in the nucleation of separated islands on the MgO substrate (Figure 1a, top panel). The surface morphology of an FeRh sample fabricated via this procedure for $t$ = 40 nm is characterized via atomic force microscopy (AFM) and shown in Figure 1a (bottom panel). The deposit consists of densely-packed sub-micron islands extending over the entire substrate. The islands display a characteristic rectangular shape with pronounced faceting along the [100] and [010] axes of FeRh (in Figure 1a, these axes are 45° rotated with respect to the image edges, corresponding to the principal axes of MgO). X-ray diffraction (XRD) measurements confirm the attainment of the CsCl-type structure and a well-defined FeRh(001) crystallographic texture (Figure S1, Supporting Information).

The formation of FeRh islands originates from surface energy optimization leading to a preferential Volmer-Weber growth mode. We find that film percolation does not occur in the first 3 min of the deposition (at a nominal film thickness of 6 nm). Ramping the growth temperature to 1100 K before forming a continuous layer results in Volmer-Weber growth due to the large surface energy difference between the deposit ($\gamma_{FeRh,100}$ = 2.17 J m$^{-2}$ [38]) and the substrate ($\gamma_{MgO,100}$ = 1.15 J m$^{-2}$ [46]). Nanoisland size analysis reveals a bimodal distribution with a broad peak at the equivalent diameter value of 260 nm and the additional presence of sub-50 nm islands (Figure S1, Supporting Information). Scanning electron microscopy (SEM) observations reveal the ubiquitous presence of these smaller nanoislands intercalated between bigger ones (>100 nm), further confirming the scenario of Volmer-Weber nucleation (Figure S2, Supporting Information).

This result is in line with previous reports of island-like growth in high-temperature-deposited (>900 K) ultrathin FeRh/MgO(001) samples [36-38]. Deposition of an FeRh film at constant, elevated temperatures above 1000 K also led to nucleation of micron-sized islands of arbitrary shapes on MgO(001) (Figure S3, Supporting Information), which we attribute to a complex balance between the temperature-dependent surface energy, deposit-to-substrate mismatch, and strain relaxation during growth and post-growth annealing.

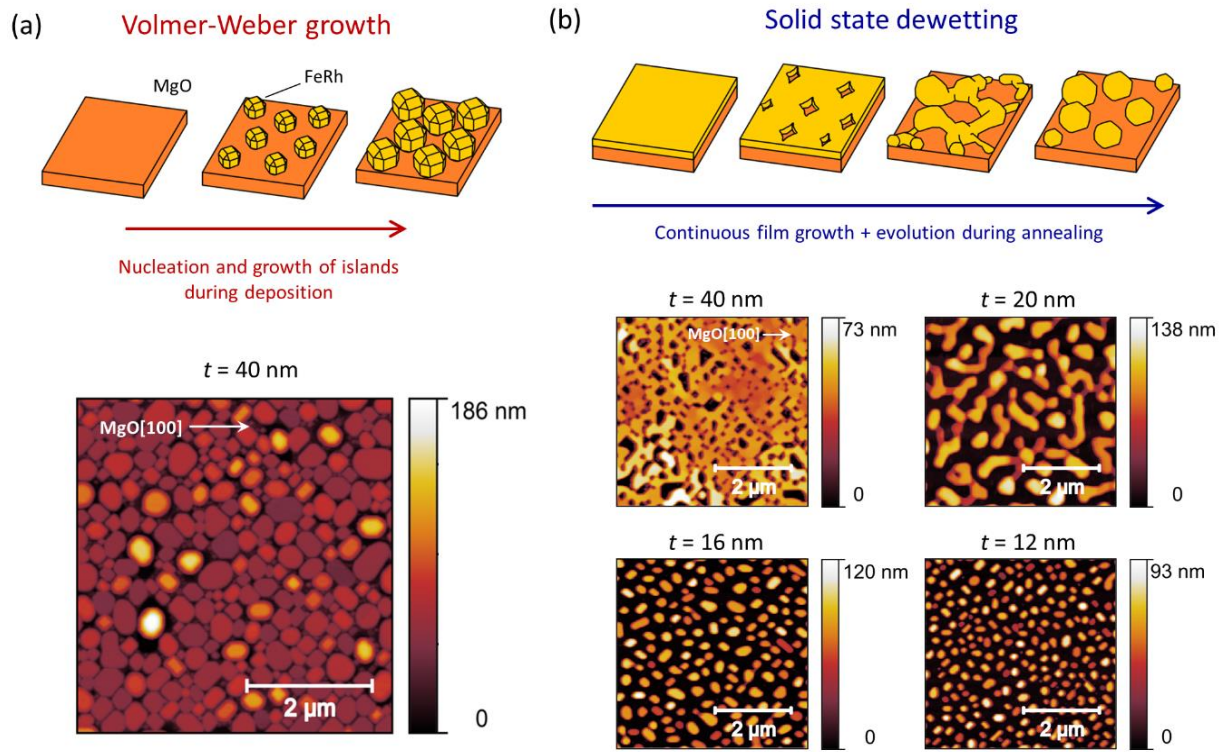

**Figure 1.** a) Schematic representation of the nanoisland self-assembly process via Volmer-Weber nucleation (top panel) and exemplary topographic height profile dataset for a sample with a nominal $t = 40$ nm thickness (bottom panel). b) Illustration of the solid-state dewetting process in FeRh thin films (top panel), and topographic height profile datasets for FeRh deposits formed via solid-state dewetting for different film thicknesses $t$ (bottom panel). As $t$ is reduced, the film morphology evolves from a continuous coverage with square-shaped voids towards well-separated, sub-micron islands of decreasing size.

The second self-assembly process, developed in this work, is initiated by the non-equilibrium growth of a continuous metastable FeRh thin film on MgO. FeRh is sputtered at a substrate temperature of 750 K throughout the entire deposition. The samples are subsequently annealed at 1100 K for 80 min. We find that the thermal load exerted during post-growth annealing is enough to drive metastable FeRh films towards the thermodynamically favored island-like morphology via solid-state dewetting [44, 45].

Spontaneous agglomeration of three-dimensional islands starts with the nucleation and deepening of grooves in the epitaxial FeRh film (Figure 1b, top panel), a process that initiates at defect sites consisting of vacancies or contaminants [44]. This step is followed by the anisotropic retraction and thickening of faceted rims around the voids, leading to hole nucleation down to the substrate [47, 48]. The occurrence of further mass transport in the form of capillary instabilities and perturbations in the fronts of the receding rims, finger formation, and Rayleigh-type instabilities [49] leads to the self-assembly of sub-micron FeRh islands (Figure 1b).

AFM images of dewetted FeRh films are shown in the bottom panel of Figure 1b. The resulting morphology is strongly dependent on the nominal film thickness. For $t = 40$ nm, FeRh/MgO samples feature an interrupted film morphology with square-shaped grooves that point towards a strongly faceted void growth along the [100] and [010] crystal axes of FeRh. Lowering $t$ leads to more advanced dewetting scenarios, in accordance with the inverse dependence of the rim retraction rate with film thickness [44, 45]. The morphology of the deposit (Figure 1b, bottom panel) evolves from maze-like, interconnected islands ($t = 20$ nm) towards physically well-separated sub-500-nm nanoislands ($t = 12, 16$ nm). We have monitored *in situ* the onset of void nucleation and growth during annealing by measuring low energy ion scattering (LEIS) on pre-grown FeRh continuous films of different thicknesses. The emergence of voids, marked by the appearance of a scattering signal from the Mg and O substrate atoms at the surface, is triggered for temperatures values of ~ 800-850 K with their growth steadily occurring up to 1100 K (Figure S4, Supporting Information). Dewetted FeRh nanoislands form arrays over the whole substrate area, with the crystallographic FeRh(001) out-of-plane texture persisting after solid-state dewetting (Figure S1, Supporting Information).

Well-separated nanoislands ($t = 12, 16$ nm) feature a characteristic oval shape with slight elongations along the principal crystal axes of FeRh, originating from the anisotropic void growth and rim faceting during dewetting. Compared to the Volmer-Weber nucleated ones, dewetted nanoislands apparently display a larger number of crystal facets, giving them a rounder geometric appearance (Figure 1b and Figure S2, Supporting Information). Furthermore, contrary to the Volmer-Weber growth mode, there is no presence of smaller intercalated nanoislands between the larger ones. Decreasing the deposited nominal thickness allows achieving nanoisland sizes down to the ~100 nm range (Figure 1b and Figure S1, Supporting Information).

As a general observation, we found that the FeRh film percolation is compromised around the nominal thickness of ~10 nm and below, identifying a crossover between solid-state dewetting and Volmer-Weber nucleation. The exact threshold strongly depends on the sample-to-sample growth temperature variations caused by the substrate-to-holder thermal contact. Hence, the lowest achievable thickness of the continuous film constitutes an intrinsic limit for decreasing the size of nanoislands formed via solid-state dewetting. Besides, the resulting size of the dewetted nanoislands depends on the density of groove nucleation sites in the initial stage of dewetting (Figure S5, Supporting Information). Groove nucleation can be triggered by pinholes in the film, impurities, dislocations, and topographical irregularities on the substrate (e.g., terraces [44]). A larger density of hole nucleation sites will typically break up the film into a larger amount of smaller nanoislands.

The detailed morphology and crystal faceting of self-assembled nanoislands were additionally analyzed on MgO(011) and Al$_2$O$_3$(0001), where we also obtained FeRh nanoisland arrays via dewetting. Dewetting of FeRh on Al$_2$O$_3$(0001) is also facilitated by the low surface energy of this surface plane, with $\gamma_{Al2O3,0001} = 1.4$ J m$^{-2}$ [50]. MgO(011) possesses a higher surface energy of $\gamma_{MgO,110} \sim 3$ J m$^{-2}$ [51], making island assembly thermodynamically unfavorable. However, moderate micro-faceting of the substrate surface lowers this value to about ~ 1.7 J m$^{-2}$ [52], hence explaining the occurrence of dewetting in our experiments.

Figure 2a shows overview AFM scans of FeRh nanoislands obtained via solid-state dewetting on MgO(001), MgO(011) and Al$_2$O$_3$(0001). The nanoislands possess high crystalline order with (001), (112) and (111) out-of-plane textures, respectively (Figure S6, S7, Supporting Information). The FeRh(001)/MgO(001) and FeRh(111)/Al$_2$O$_3$(0001) epitaxial matchings have already been described for continuous films [18, 19], whereas we find that the epitaxy of FeRh on MgO(011) follows a pattern common for elemental bcc metals like Fe and Cr, showing (112)-oriented preferential growth on MgO(011) [53, 54].

Anisotropic rim retraction following void formation during solid-state dewetting strongly determines the shape of self-assembled nanoislands. Topographic characterization of FeRh films during early dewetting stages (Figure S8, Supporting Information) hints to nanoisland arrangements with characteristic four-fold, two-fold and six-fold symmetries for the (001), (112) and (111) oriented cases, respectively. FeRh(001) nanoislands elongate along the [100] and [010] axes of FeRh (Figure 2a). Further, FeRh(112) nanoislands predominantly elongate along the FeRh[11$\underline{1}$]||MgO[0$\underline{1}$1] direction (vertical direction in Figure 2a), sometimes forming high aspect-ratio needles joining the nanoislands. Finally, FeRh(111) nanoislands possess slightly elongated oval shapes aligned along orientations distributed every 60° in the plane. For the latter two substrates, we typically observe more densely packed and smaller nanoislands than those obtained on MgO(001), as a larger density of hole nucleation sites arises from the higher FeRh-to-substrate epitaxial mismatch and the subsequent larger presence of stacking faults and dislocations (Figure S9, Supporting Information).

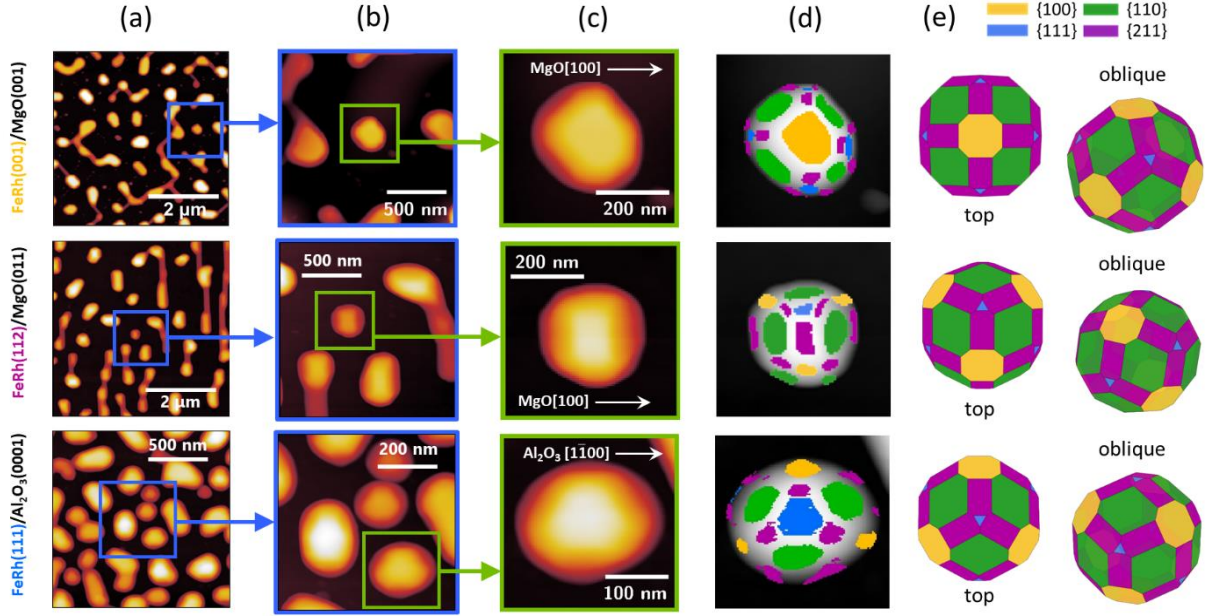

**Figure 2.** Morphology and faceting of self-assembled epitaxial FeRh nanoislands on single-crystal oxide substrates. a) Topographic AFM data with tens of nanoislands for samples with FeRh(001), FeRh(112) and FeRh(111) out-of-plane crystallographic orientations ($t = 16$ nm). b) and c) show zoomed-in topographic scans showing a few nanoislands and a single nanoisland, respectively. d) Facet analysis obtained from the high-resolution, single-nanoisland topography scan. e) Top and oblique views of the Wulff construction for each of the different nanoisland textures.

The shape of selected nanoislands ~200 nm in size has been studied in detail for the distinct out-of-plane crystallographic orientations (zoomed-in AFM images in Figure 2b, c). The analysis of surface normal orientation distributions in high-resolution topographic scans allows crystallographic facet identification (see Experimental Section). Figure 2d shows the marked crystal facets superimposed with topography, side-by-side to the bare data in Figure 2c, upon considering the {100}, {110}, {111}, and {211} crystal facets of FeRh.

We have modelled the equilibrium crystal shapes or Wulff constructions (see Experimental Section) of FeRh crystals using the surface energies calculated by Liu *et al.*, with $\gamma_{100} = 2.17$ J m$^{-2}$, $\gamma_{110} = 2.10$ J m$^{-2}$ and $\gamma_{111} = 2.37$ J m$^{-2}$ [38]. In addition, we set $\gamma_{211} = 2.20$ J m$^{-2}$ to match the topography line profiles to the equilibrium crystal shapes (Figure S10, Supporting Information). The modelled equilibrium shapes for each nanoisland texture are shown in Figure 2e. The experimentally determined nanoisland faceting and equilibrium shapes (Figure 2d, e) agree very well, indicating that dewetting leads to the assembly of equilibrium FeRh crystal shapes. The facet analysis also confirms the epitaxial matching relations deduced from XRD (Figure S7, Supporting Information).

## 2.2 Magnetic properties of FeRh nanoislands

The magnetic properties of the FeRh nanoislands assembled via solid-state dewetting are first analyzed at room temperature. Figure 3a-c show AFM micrographs of a 5 × 5 μm$^2$ area containing sub-micron nanoislands with (001) out-of-plane texture and different nominal thicknesses. The room-temperature magnetic force microscopy (MFM) measurements in Figure 3d-f correspond to the topographic scans displayed above. The magnetic signal principally arises from out-of-plane oriented magnetic moments due to the externally applied vertical magnetic field (see Experimental Section). The morphology of islands with $t$ = 20 nm (Figure 3a) corresponds to a maze-like structure with well-recognizable dewetted features. The magnetic signal (Figure 3d) is only present in three distinct regions, which disrupt the otherwise prevailing zero contrast background (represented by the green color in Figure 3d). The major fraction of FeRh islands thus manifests zero magnetic signal at room temperature.

Nanoislands corresponding to $t$ = 16 nm show similar features compared to the sample with $t$ = 20 nm, but the fraction of islands showing magnetic contrast is larger (Figure 3b, 3e). About half of the well-separated islands exhibit zero MFM signal, with the remaining half revealing a clear FM ordering (Figure 3e). The relative population of islands showing magnetic signal is even larger for the sample with $t$ = 12 nm, containing well separated ~200 nm nanoislands (Figure 3c, 3f), and where only a few islands show zero magnetic signal. This island-size dependent analysis thus reveals that with decreasing size, a larger fraction of nanoislands displays a clear FM ordering at room temperature. The magnetic properties of FeRh nanoislands with (112) and (111) textures were also evaluated. MFM measurements indicate that almost all FeRh(112) nanoislands remain FM at room temperature. In case of FeRh(111), about half of the ~100 nm nanoislands show a significant MFM signal (Figure S9, Supporting Information)

In order to characterize the AF-FM phase transition in the nanoisland samples, the temperature dependence of magnetization was measured using vibrating sample magnetometry (VSM). The magnetization data are shown for each sample in the panels below the corresponding topographic and MFM data (Figure 3g-i). Apparently, all FeRh(001) nanoisland samples undergo a prominent phase transition. Overall, the heating cycle of the thermal hysteresis shows an abrupt phase transition, while the magnetization drop during the cooling cycle is more gradual upon decreasing the size of the nanoislands. For larger nanoislands, the phase transition during cooling is abrupt and only shows a slight tail marking the need for a cool-down slightly below room temperature in order to complete the transition ($t$ = 20 nm, Figure 3g).

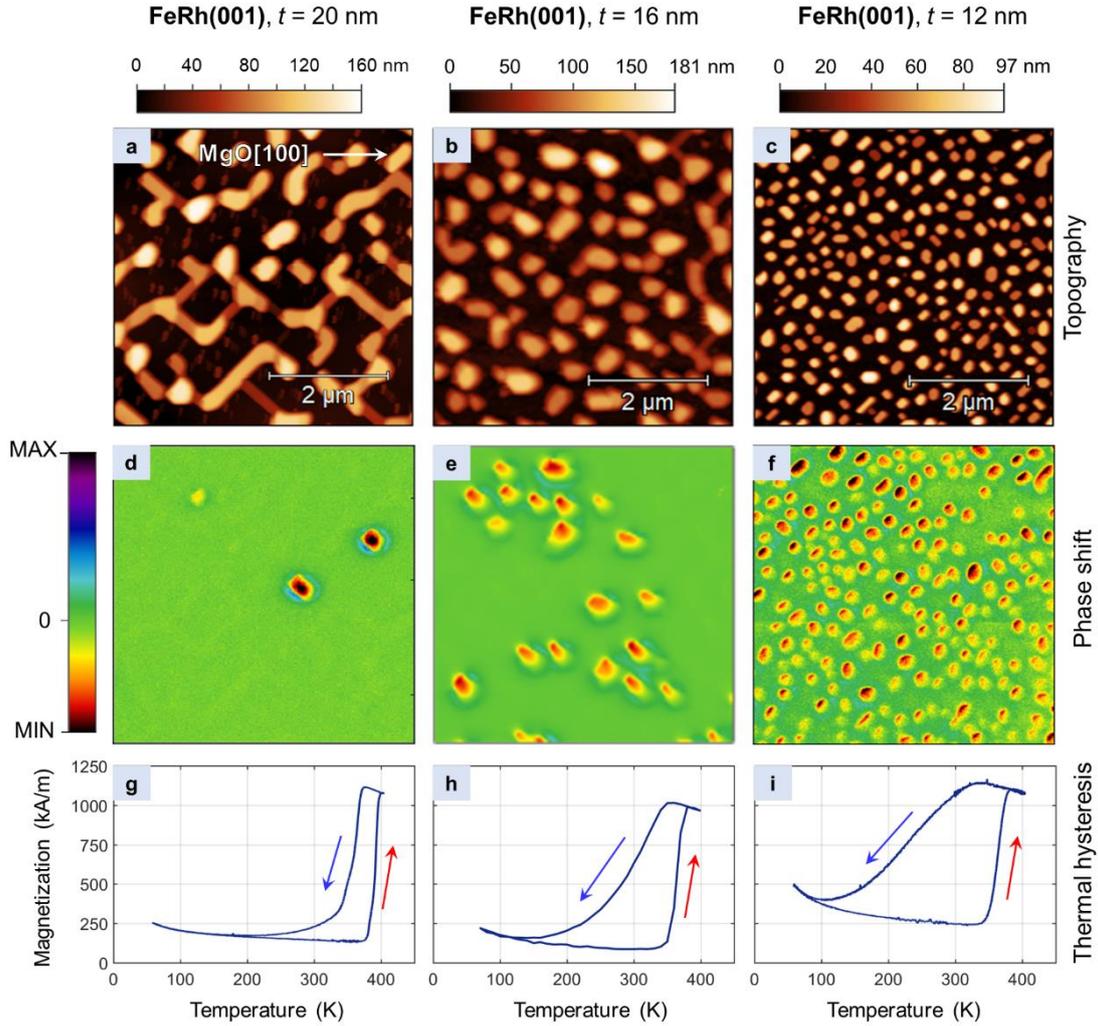

**Figure 3.** Room temperature magnetic properties and phase transition in FeRh nanoislands. a)-c) AFM topography images over a 5 × 5 μm² area of FeRh nanoisland samples with $t$ = 20, 16, 12 nm. d)-f) Room-temperature MFM measurements over the same sample area. The inset in a) indicates the crystallographic in-plane direction of the micrograph, which is also valid for all panels b)-f). g)-i) Temperature dependence of the magnetization in the range 55-400 K for the samples described above. The arrows indicate the heating and cooling cycles in the thermal hysteresis.

As the nanoislands' size decreases, the thermal hysteresis features a more gradual change and a broad transition of magnetization during cooling ($t$ = 16, 12 nm, Figure 3h, i). For these small nanoislands, a considerable fraction of the high-temperature magnetization is retained at room temperature during cooling, and the phase transition is only completed after cooling down the sample below 150 or 100 K. This observation agrees well with the large fraction of nanoislands showing magnetic MFM signal at room temperature (Figure 3e, f). We conclude that a large fraction of nanoislands with sizes around and below 200 nm remain supercooled in the FM phase at room temperature. This behavior is also found in the case of

(112) and (111)-textured nanoislands, where a prominent phase transition is equally present (Figure S9, Supporting Information). Finally, we have investigated the magnetic field dependent magnetization behavior of FeRh(001) textured nanoislands (Figure S11, Supporting Information). Field hysteresis loops measured at different temperatures indicate a slight out-of-plane preferential magnetization orientation, as predicted in the literature [55].

In the following, we present the magnetic behavior of individual dewetted nanoislands across the phase transition. Figure 4a shows the topography of FeRh(001) islands ($t = 16$ nm) over an $8 \times 8$ µm$^2$ area. As the nanoislands are firstly heated across the AF-to-FM phase transition, it is seen that most of them become FM within the temperature range from 343 to 368 K, as evidenced by a clear magnetic signal in the MFM scan (Figure 4b-d).

The cooling characteristics is investigated over a larger $10 \times 10$ µm$^2$ sample area (Figure 4e). Here, we combine *ex-situ* cooling (down to 100 K) and heating (up to 400 K) of the samples with posterior MFM observation at room temperature. Figure 4f shows temperature-dependent magnetization data for the different heating/cooling protocols performed prior to the room-temperature MFM characterization. The initial magnetic configuration of the islands at room temperature is shown in Figure 4g. Apparently, a certain fraction of the islands is FM-stabilized at room temperature. The sample is subsequently warmed up to 400 K, followed by a cool-down to 250 K. Imaging the magnetic configuration at room temperature after this protocol reveals that a number of previously islands in the FM phase show no magnetic signal (Figure 4h), suggesting that these islands were supercooled at room temperature and underwent the FM-to-AF phase transition upon additional cooling down to 250 K. The temperature protocol and imaging are repeated while decreasing the cooling temperature to 200 K (Figure 4i), 150 K (Figure 4j) and 100 K (Figure 4k). We observe that the number of FM-stabilized FeRh nanoislands decreases upon each consecutive cooling protocol. After cooling down to 150 K, only three nanoislands remain FM (Figure 4j), and finally, we find that all nanoislands have transitioned to the AF phase upon cooling down to 100 K (Figure 4k).

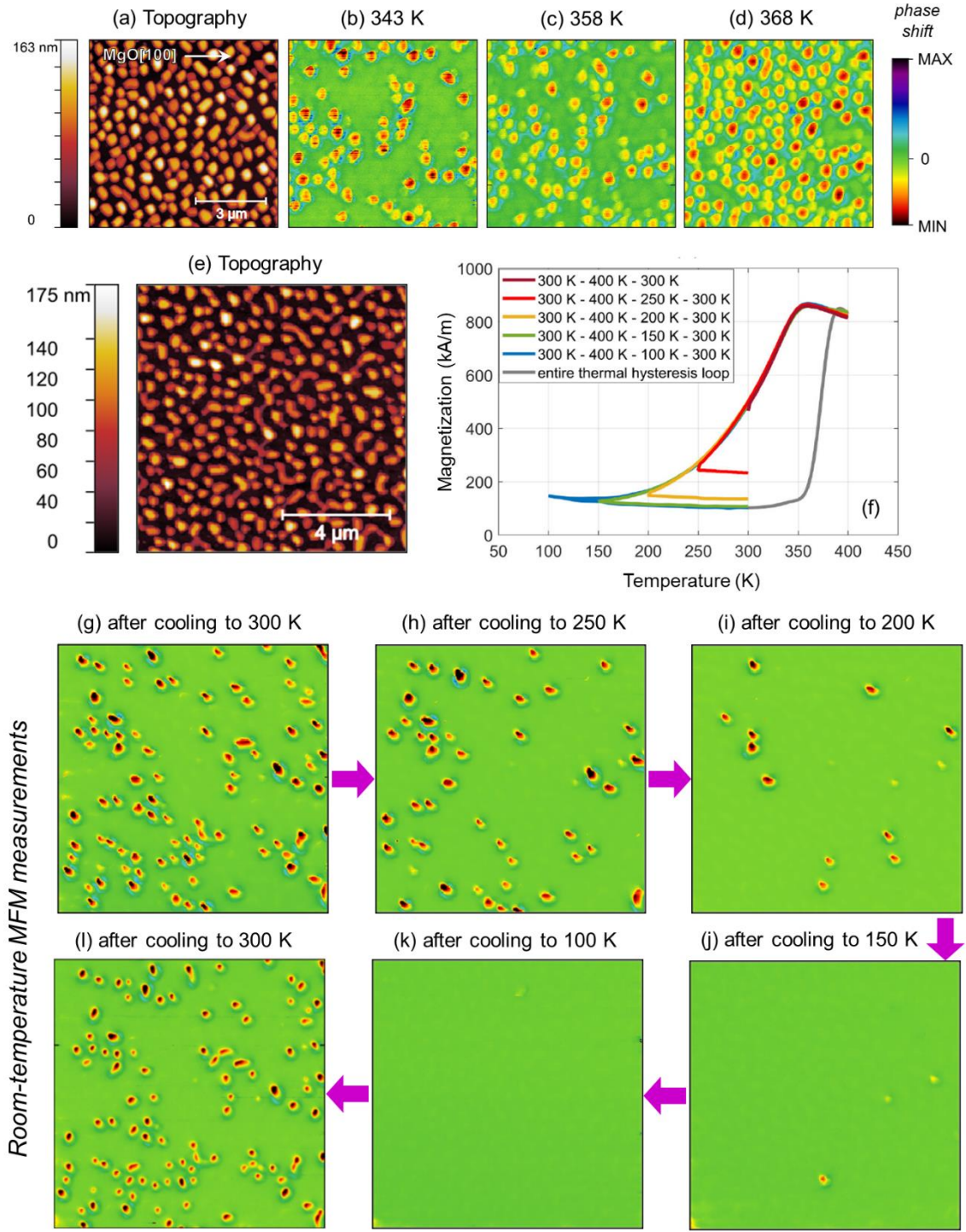

**Figure 4.** Phase transition and supercooling in FeRh nanoislands ($t = 16$ nm). a) AFM image over an $8 \times 8$ um$^2$ area and b)-d) MFM images of FeRh(001) nanoislands during heating. e) AFM image of over a $10 \times 10$ μm$^2$ area. f) Magnetization vs temperature for the sequential thermal cycling employed before room temperature MFM characterization. g)-l) room-temperature MFM measurements upon warming up the sample to 400 K and subsequently cooling down to the temperatures indicated in each panel. The arrows between panels g)-l) indicate the order of the sequential heating and cooling cycling.

The supercooled nanoislands transition to the FM phase well above 300 K, in the range ~350-370 K regardless of the thermal cycling protocol. This indicates that sub-micron FeRh nanoislands can present very extensive supercooling at about 150-200 K below their transition temperature (for comparison, the deep supercooling regime for the liquid-to-ice phase transition in water is ~43 K below the freezing point [56]). While supercooling of about ~10-20 K has been previously reported in lithographically patterned FeRh wires [57, 58], we observe that self-assembled FeRh nanoislands are capable of sustaining much deeper supercooled FM states. Finally, the sample is warmed up to 400 K and cooled down to 300 K for obtaining an additional snapshot of its magnetic state (Figure 4l). The magnetic order of the nanoislands closely resembles that of the initial state at 300 K (Figure 4g), but a few additional islands seem to be in the FM state.

These findings altogether point to the extraordinary sensitivity of the phase transition in confined FeRh structures to factors such as defects, availability of nucleation sites, and thermal activation. In particular, the decrease in the number of AF phase nucleation sites upon reducing the size of FeRh nanomagnets seems to be behind the observation of the very pronounced supercooling in the well-ordered nanoislands investigated here. This aspect could also be at the origin of the complete suppression of the phase transition in FeRh at the ~10 nm scale and below, as observed in highly ordered nanoparticles embedded in a carbon matrix, where the FM phase persists down to 2 K [59, 60].

Another interesting observation in the FeRh nanoislands is the complete suppression of phase separation across the phase transition. We did not observe any coexistence of AF and FM domains upon temperature cycling for the nanoisland samples presented here, indicating that the abrupt nature of the first-order phase transition is recovered within each nanoisland.

The emergence of asymmetric thermal magnetization hysteresis, with a relatively abrupt transition during heating and broad transition during cooling, has been frequently observed in FeRh specimens in the literature, particularly in thin films [61-65]. As (ultra)thin films of FeRh grown on single-crystal oxide substrates often turn out to be discontinuous or granular, we suggest that supercooling is responsible for the occurrence of this phenomenon, where a substantial presence of supercooled nanoscale grains within FeRh films could explain the appearance of such asymmetric thermal hysteresis. It is worth noting that engineering the sputter process can improve to a certain degree the continuity and smoothness of ultrathin FeRh films on oxide substrates [66].

## 2.3 Morphology enabled or hindered phase transition

While the metamagnetic behavior is preserved in the dewetted FeRh nanoislands with lateral sizes of ~300 nm and below, the phase transition is strongly suppressed in Volmer-Weber nucleated nanoislands of similar or larger size (Figure 5a). First-principles calculations by Liu *et al.* point to a strong link between a given magnetic phase and the surface energy of the principal FeRh crystal facets [38]. The minimum surface energy, and thus preferential faceting orientation, is predicted for the {110} planes in the AF phase, whereas the {100} planes are the ones with lowest surface energy in the FM phase (see Table 1).

**Table 1.** Surface energy values for the main facets of FeRh.

| FeRh plane | $\gamma$ [J m$^{-2}$] AF phase[a] | $\gamma$ [J m$^{-2}$] FM phase | Ref. |
|:---:|:---:|:---:|:---:|
| {100} | 2.17 | 1.78 | [38][a] |
| {110} | 2.10 | 1.87 | [38][a] |
| {111} | 2.37 | 2.08 | [38][a] |
| {211} | 2.20 | -- | This work[b] |

[a] The values from Liu *et al.* are calculated by density functional theory for both AF and FM bulk phases [38]. We assume a Rh-terminated surface for each case and FM surface configurations in the AF phase [40-42]; [b] The energy for the {211} planes is obtained by matching the experimental topographic profile while setting the theoretical surface energy values for the remaining facets in the profile (Figure S10, Supporting Information).

Figure 5b shows AFM scans for selected nanoislands in the Volmer-Weber nucleated and dewetted samples, respectively. Both nanoislands are similar in lateral size, but their morphology is qualitatively different. Apparently, the island assembled via Volmer-Weber nucleation shows a prevailing {100} crystal faceting with a characteristic rectangular shape, while the dewetted island has a rounder morphology arising from the predominant {110} crystal plane faceting. Topographic line scans along the FeRh[110] direction for the these two nanoislands (Figure 5c) reveal that while both nanoislands correspond to nanocrystals truncated above their center. However, while the Volmer-Weber nucleated island features a relatively low height-to-diameter ratio and a prominent faceting for the {100} planes, the dewetted island has a noticeably larger height in proportion, with a larger relative presence of the {110} facets.

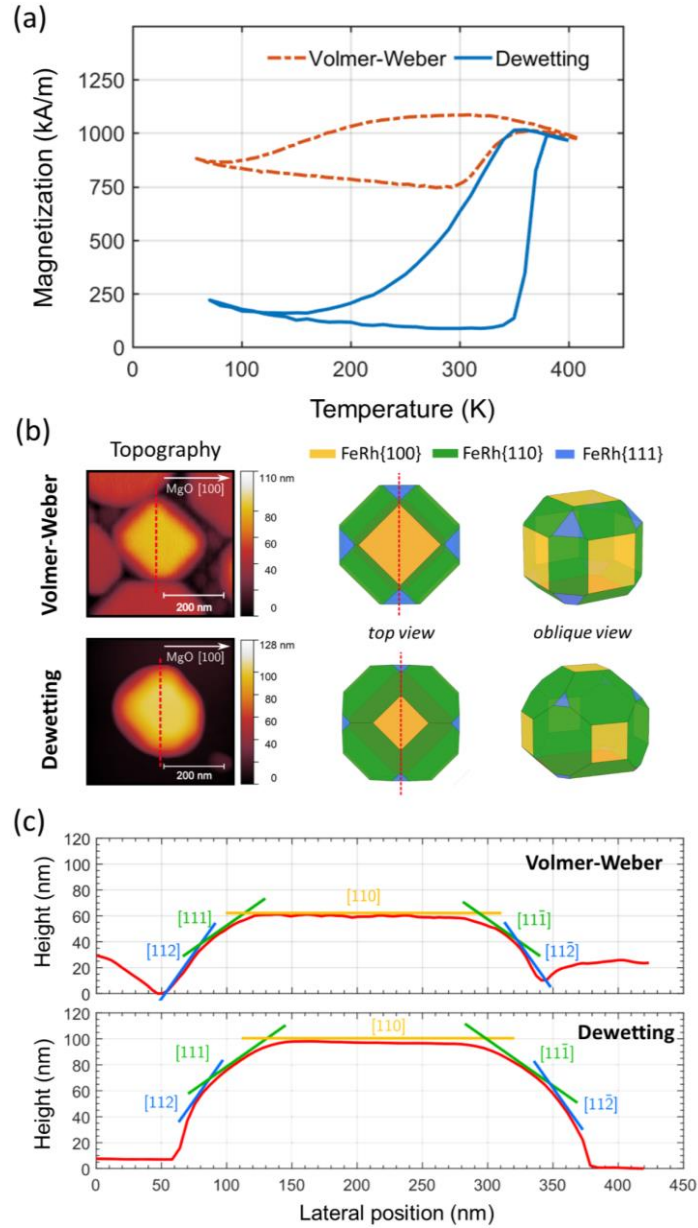

**Figure 5.** Morphology and metamagnetism in FeRh nanoislands. a) Magnetization vs temperature measurements for FeRh(001)/MgO(001) nanoislands assembled via Volmer-Weber growth ($t = 40$ nm) and solid-state dewetting ($t = 16$ nm). b) Topography scans showing a single nanoisland of ~300 nm in diameter for both Volmer-Weber growth and solid-state dewetting. Wulff constructions of FeRh nanocrystals obtained upon considering the surface energies of the FM and AF phase, are shown side-by-side to the topography scans, which resemble Volmer-Weber nucleated and dewetted nanoisland morphologies, respectively. The representation of {211} planes has been omitted for clarity. c) Topography line scans of Volmer-Weber and dewetted single nanoislands extracted from the data in b). The indices within brackets in c) denote the crystallographic directions along the line-scan for each facet.

It is interesting to notice that the Wulff nanocrystal models obtained by choosing the surface energy values for the AF or FM phases (Table 1) qualitatively predict the experimental nanoisland shapes (Figure 5b). That is, equilibrium FM FeRh nanocrystals resemble the morphology of FeRh nanoislands obtained via Volmer-Weber nucleation, while AF FeRh nanocrystals resemble nanoislands assembled via solid-state dewetting (Figure 5b).

In order to further elucidate the contrasting magnetic behavior of nanoislands assembled via different routes, we have thoroughly analyzed the topographic features of nanoisland ensembles on MgO(001) substrates formed via Volmer-Weber growth and solid-state dewetting. Two characteristic length scales are measured from each island or truncated crystal: the base to cusp height $h$ and the extent of the cusp $L$ in the FeRh[110] direction (see Figure 6a). We perform a statistical analysis of the $h/L$ ratio by considering 75 islands from each sample (Figure 6b). The statistical analysis of this morphological feature confirms the differences anticipated in Figure 5c for the two types of nanoislands. The central value obtained from the $h/L$ histogram is $0.3 \pm 0.1$ for Volmer-Weber nucleated islands and $0.8 \pm 0.2$ for those ones assembled via solid-state dewetting, pointing to a marked difference in the morphology of nanoislands assembled via each route.

Next, we employ the Wulff-Kaischev's theorem, which mathematically relates the occurrence and geometry of the facets in a supported crystal with the surface energy values of the crystal and the substrate, as well as the with interface formation energy [67, 68]. Following this approach, we arrive to the expression (Note S1, Supporting Information),

$$\gamma_{\text{int}} = (\gamma_S - \gamma_{100}) + (4\gamma_{110} - 2\sqrt{2}\gamma_{100})\frac{h}{L}$$

(1)

which relates the FeRh/MgO interface energy to the measured nanoisland $h/L$ ratio and surface energies of FeRh and the MgO substrate. Figure 6c shows the dependence of the interface energy on $h/L$ according to Equation 1, for the cases in which the AF and FM surface energy values are considered. Negative $\gamma_{\text{int}}$ values do not represent an accessible physical solution, values with $\gamma_{\text{int}} > \gamma_S \approx 1.17$ J m$^{-2}$ correspond to nanocrystals truncated below their geometric center, which were not experimentally observed.

Using the central $h/L$ values in the histogram for the two types of islands, we observe that for a morphology corresponding to that of Volmer-Weber nucleated islands, the only allowed interface energy exists upon assuming FeRh surface energy values in the FM phase ($h/L = 0.3$, $\gamma_{\text{int}} = 0.17$ J m$^{-2}$), while for dewetted nanoislands both phases are accessible, with the AF phase representing the more stable configuration ($h/L = 0.8$, $\gamma_{int} = 0.47$ J m$^{-2}$) and the only one corresponding to a truncation above the nanocrystal center.

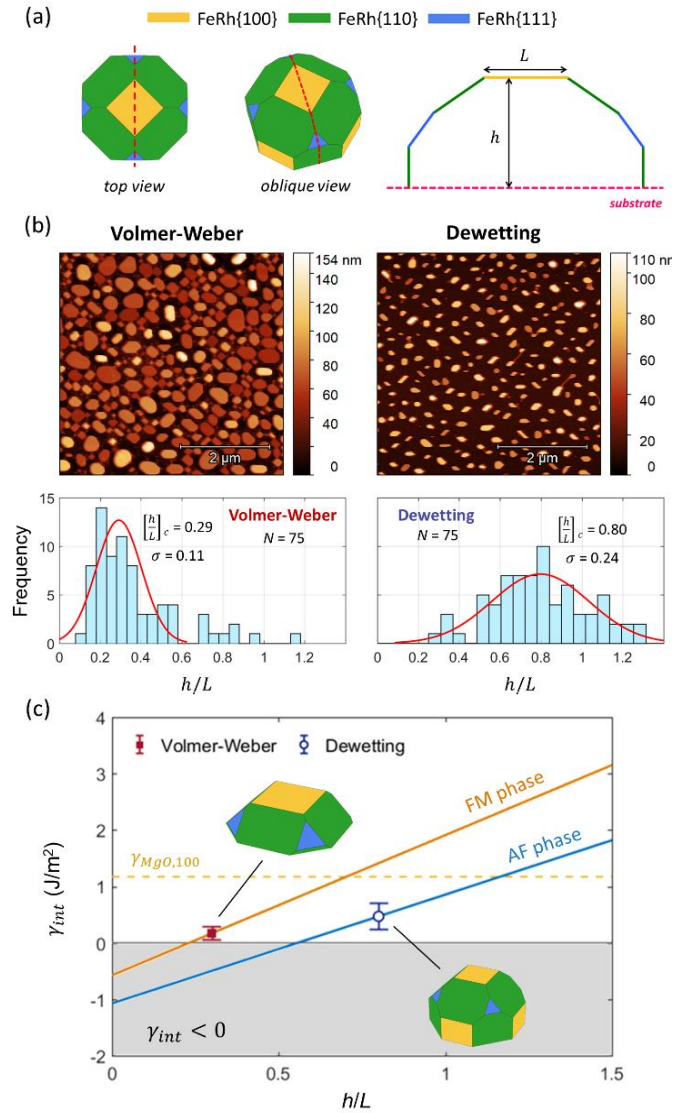

**Figure 6.** Shape analysis of FeRh nanoislands. a) Top and oblique view of the Winterbottom construction for a (100)-oriented nanoisland truncated above the nanocrystal center. The {211} planes are omitted for clarity. The red dashed lines indicate the line-scan orientation measured for each island. On the right, schematics of the nanoisland attributes (height $h$, cusp width $L$) extracted from the line-scans. b) High-resolution AFM image (5 × 5 μm$^2$) and histograms of the measured $h/L$ ratio for $N = 75$ nanoislands obtained via Volmer-Weber nucleation and dewetting. The fitted central $h/L$ value and the standard deviation are indicated (for Volmer-Weber nanoislands, outliers with $h/L > 0.6$ are neglected). c) FeRh/MgO interface energy vs $h/L$ considering the AF and FM phases of FeRh (solid lines), and values obtained from the measured $h/L$ distributions for Volmer-Weber nucleated and dewetted nanoislands. The schematics in the inset in c) show the shapes of the truncated nanocrystals in each case.

The manifest differences in the shape and magnetic phase transition properties of the FeRh nanoislands formed via different assembly routes point to a very strong connection between their nanocrystal morphology and the favored magnetic order. Our study provides strong evidence for this connection, as we can conclude that FeRh nanoislands with distinctive shapes tend to sustain or preclude the AF phase, and in turn metamagnetism, in a scenario that is compatible with the phase-dependent calculated surface energies of FeRh [38].

**2.4 Free-standing FeRh nanoparticles**

We released the supported FeRh(001) nanoislands on MgO(001) from the substrate in order to study their magnetic properties as free-standing nanoparticles. Figure 7a shows an AFM image of FeRh nanoislands assembled from a 12-nm-thick film via solid-state dewetting, featuring typical sizes of 200 nm and below. Nanoislands were separated from the substrate via chemical etching of MgO (see Experimental Section).

Figure 7b shows the field hysteresis at 400 K for the FeRh nanoislands before and after being released from the substrate. The measured maximum magnetic moment of 0.32 $\mu Am^2$ for the supported islands agrees well with that of a nominally 12-nm-thick film with a magnetization value of 1120 kA m$^{-1}$ [18, 19]. Likewise, the maximum magnetic moment value of 0.23 $\mu Am^2$ measured for the released nanoparticles allows estimating that ~72% of the nanoislands were recovered. Considering the nanoparticles as platelets with an average nanoparticle diameter and thickness of 200 and 60 nm, respectively (Figure S12, Supporting Information), it can be estimated that ~$1.2 \cdot 10^8$ nanoparticles were obtained after separation.

The temperature-dependent scaled magnetic moment is shown in Figure 7c for both nanoislands and nanoparticles. The phase transition characteristics are very similar for the FeRh nanomagnets when supported and released from the substrate, with a virtually identical temperature dependence of the magnetic moment for the heating and cooling cycles. As in the case of supported nanoislands, free-standing nanoparticles show a relatively abrupt increase of the moment during heating and a smoother decrease during cooling (Figure 7c), thus the prominent supercooling behavior being kept after release from the substrate. It is concluded that the substrate-induced strain in ~200 nm size nanoislands is largely relaxed as a result of the surface-to-volume ratio increase upon dewetting, opposite to continuous thin films where detachment from the substrate causes large shifts of the phase transition temperature [62].

We find a slight difference for supported and free-standing FeRh nanomagnets in terms of the residual FM moment fraction in the nominal AF phase (see heating cycle in Figure 7c), where it is about ~ 20% higher in the latter case. We explain this in terms of the nanoparticle separation process, where nanoparticles in the FM phase were likely captured more efficiently than those in the AF state. The ~28% fraction of non-recovered nanoparticles could show a comparatively lower amount of residual magnetic moment, explaining the difference. Another possibility is that nanoparticle accumulation could stabilize the FM phase within these clusters, thus suppressing AF order within a limited fraction of nanoparticles.

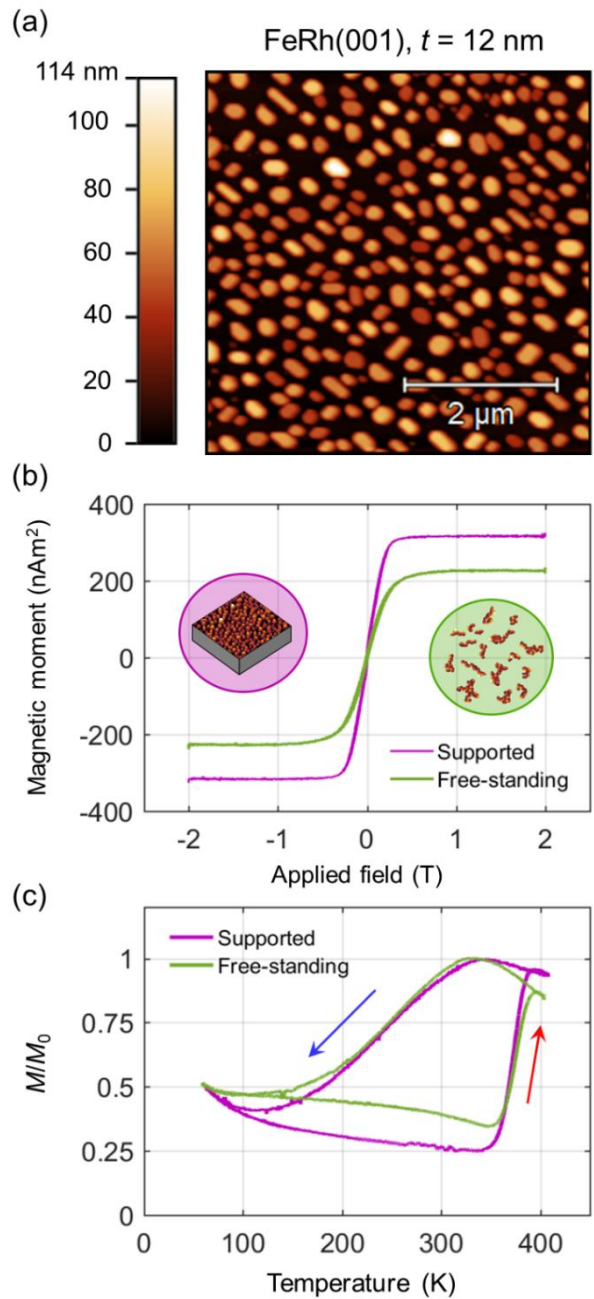

**Figure 7.** Metamagnetism in free-standing FeRh nanoparticles. a) AFM image of FeRh(001) islands before etching. b) Magnetization vs applied field for the supported islands and the released nanoparticles at 400 K. The inset in b) shows schematics of the supported and free-standing FeRh nanomagnets. c) Scaled magnetic moment vs temperature at 1 T for the supported and detached FeRh islands. The red and blue arrows indicate the heating and cooling cycles, respectively.

## 3. Conclusions

In conclusion, we have investigated self-organization of metamagnetic FeRh nanoislands using sputter deposition. Two different routes lead to the self-assembly of epitaxial and sub-micron nanomagnet arrays on single-crystal oxide substrates. On the one hand, Volmer-Weber nucleation leads to densely-packed nanoislands with predominant faceting along the principal axes of FeRh. On the other hand, growth of a metastable continuous film and subsequent solid-state dewetting leads to multifaceted nanoislands with sizes down to ~100 nm. The size and shape of nanoislands assembled via dewetting can be controlled via epitaxy and by tuning the deposited FeRh thickness. While we find that the phase transition is strongly suppressed in sub-micron islands nucleated during Volmer-Weber growth, dewetted FeRh islands show largely preserved metamagnetism.

Tracking the magnetic properties of a single nanoisland upon temperature cycling reveals large confinement effects such as very pronounced supercooling (>150 K) and the absence of phase separation in sub-500-nm nanoislands. The detailed comparison of the specific crystal faceting and magnetic properties of nanoislands assembled via nucleation and dewetting permits establishing that nanoscale morphology has a strong impact on the phase transition characteristics of nanoscale FeRh systems. We find that nanoislands showing a predominant {100} crystal faceting are strongly FM stabilized, while those showing a prevailing {110} faceting can undergo the phase transition to the AF phase.

Self-assembly of FeRh islands on oxide substrates could be further controlled via templated dewetting by making use of pre-patterned substrates or films [69-71], optimizing aspects such as the nanoisland lateral size distribution or enabling the fabrication of regularly spaced arrays.

Finally, we have also released metamagnetic FeRh nanoislands from the substrate using chemical etching and have studied the phase transition characteristics of self-standing nanoparticles. The magnetic properties of the released FeRh islands do not significantly vary with respect to the supported case, and exhibit almost identical phase transition temperatures, supercooling behavior, and residual fraction of magnetic moment. We envision the possibility to fabricate more substantial amounts of functional FeRh nanoparticles via sputter deposition and solid-state dewetting on larger area substrates. Based on nanoisland densities of ~ 5-10 $\mu m^{-2}$ and considering the typical nanoisland height and lateral size values obtained here, the utilization of larger scale wafers [72] would enable producing ~ $10^{10}$ - $10^{11}$ nanoparticles by using, e.g., 4-in. (102 mm) wafers, thus reaching milligram mass ranges of metamagnetic FeRh in the form of nanoparticle ensembles.

## 4. Experimental Section

**Sample growth and self-assembly:** FeRh thin films were sputter-deposited onto single-crystal MgO(001), MgO(011), and $Al_2O_3$(0001) substrates (5×5 mm$^2$ in size) from an equiatomic FeRh target in a high-vacuum chamber with a base pressure of $5 \times 10^{-8}$ mbar. All substrates were preheated to 750 K in high vacuum for 1 hour in order to outgas and reconstruct the oxide surface. Unless otherwise indicated, FeRh growth was performed at a substrate temperature of 750 K and an Ar pressure of $2.7 \times 10^{-3}$ mbar. The deposition rate for FeRh was calibrated via x-ray reflectivity for continuous films and determined to be 2 nm min$^{-1}$. To fabricate the FeRh nanoislands, thin-films were post-growth annealed *in situ* and in high vacuum at 1100 K for 80 min to induce self-assembly via solid-state dewetting, as well as to improve the bcc-like structural and chemical ordering. The samples were taken out to air after they were cooled down below 373 K.

**In-situ surface elemental analysis:** The elemental composition of the sample surface during solid-state dewetting was monitored during the course of post-growth annealing using LEIS in uncapped FeRh thin films that were previously sputter-deposited in a separate chamber. The extreme surface sensitivity of LEIS allows providing straightforward identification of elements in the outermost surface layer, with the measured signal intensities reflecting the surface concentration of the detected elements [73, 74]. We have used a 3 keV He ion beam at a scattering angle of 145° to obtain a spectrum of the sample while steadily ramping up the temperature (9 K min$^{-1}$) from 300 to 1100 K. The atomic mass of the target atoms can be obtained by tracking the kinetic energy of the He projectile and following the rules of elastic binary collisions [75].

**Atomic and magnetic force microscopy:** AFM/MFM measurements were realized using a Dimension Icon microscope from Bruker Corporation. The majority of the data were acquired by employing commercial MESP probes with a hard magnetic CoCr coating. Their resonance frequency is about 75 kHz and the spring constant amounts to 3 N m$^{-1}$. Topography (AFM) was acquired in the *PeakForce Tapping* non-resonant mode which responds to short-range interactions. MFM is measured in 2$^{nd}$ pass (interleave) *LiftMode* via monitoring the phase shift of the oscillating cantilever driven near its resonant frequency. MFM images are acquired in a constant external magnetic field of ~0.3 T provided by a permanent magnet. The field is applied in out-of-plane direction to facilitate visualization of the ferromagnetic phase. High-resolution AFM images were acquired using Olympus OMCL-AC240TS probes with a nominal tip radius of 7 nm, a cantilever resonance frequency of 70 kHz, and a spring constant of 2 N m$^{-1}$. Temperature control during AFM/MFM measurements is achieved via a custom-made sample holder based on Peltier modules and providing a regulation in the range of 290-380 K at ambient conditions. AFM/MFM data were analyzed and visualized using the open-source *Gwyddion* software [76]. The modelling and depiction of individual nanoisland morphologies were realized using the *WulffPack* Python package [77], which enables the prediction of the Wulff and Winterbottom constructions of a given nanocrystal provided its crystallographic structure, facet-dependent surface energies and the nanocrystal/substrate interfacial formation energy. The analysis of experimental nanoisland morphologies and crystallographic facet determination was performed with the assistance of the in-built facet analysis tool in *Gwyddion*, in combination with modeling in *WulffPack*.

**Structural analysis:** XRD measurements were performed using a Rigaku Smartlab 9 kW diffractometer with Cu-K$_\alpha$ radiation ($\lambda$ = 1.5406 Å) using a double-bounce Ge(022) monochromator and a 5° Soller slit in the incident and diffractive optics, respectively.

**Electron microscopy imaging:** SEM images were acquired using a high resolution Verios 460L microscope by FEI, using indistinctively secondary electrons or backscattered electrons.

**Magnetization measurements:** Temperature dependent magnetization measurements were performed via VSM using a Quantum Design Versalab magnetometer in the temperature range 55-400 K and under an in-plane applied magnetic field of 1 T. Field hysteresis loops at different temperatures were recorded for in-plane or out-of-plane applied field configurations. The magnetization of nanoisland samples was calculated assuming an FeRh volume equivalent to a film with the deposited nominal thickness. All data are presented after subtracting the diamagnetic substrate contribution.

**Etching and separation of FeRh nanoislands from the substrate:** FeRh nanoislands supported on MgO(001) substrates were released in a 0.3 M solution of the disodium salt of ethylenediaminetetraacetic acid (EDTA), which was reported effective to etch MgO substrates (rate ~ 0.8 μm h$^{-1}$) and release continuous metallic films [78]. Ultrasonication did not produce any visible nanoisland detachment [79], most likely due to the strong epitaxial clamping to the substrate. The required amount of EDTA disodium salt for a 0.3 M solution was dissolved with the aid of a magnetic stirrer at 363 K to speed up the process. Subsequently, MgO(001) substrates with the fabricated FeRh nanoislands on top were inserted in the solution and kept in an oven at 348 K for ~ 30-90 min, until producing the release of the majority of nanoislands from the substrate. The released FeRh nanoparticles were separated from the EDTA disodium salt solution using a magnetic separation procedure and collected in a polypropylene capsule suitable for VSM measurements (Figure S13, Supporting Information).

## Acknowledgments

We thank Olivier Fruchart for fruitful discussions and Jiří Liška for assistance with electron microscopy. Access to the CEITEC Nano Research Infrastructure was supported by the Ministry of Education, Youth and Sports (MEYS) of the Czech Republic under the project CzechNanoLab (LM2018110). L.M. was supported by the student scholarship of Thermo Fischer Scientific. This work has received funding from the European Union's Horizon 2020 research and innovation program under the Marie Skłodowska-Curie and it is co-financed by the South Moravian Region under grant agreement No. 665860.

## Conflict of Interest

The authors declare no conflict of interest.

Supporting Information

# Preserving Metamagnetism in Self-Assembled FeRh Nanomagnets

Lucie Motyčková[1,2], Jon Ander Arregi[1,*], Michal Staňo[1], Stanislav Průša[1,2], Klára Částková[1,3], and Vojtěch Uhlíř[1,2,*]

[1]*CEITEC BUT, Brno University of Technology, Purkyňova 123, 612 00 Brno, Czech Republic*

[2]*Institute of Physical Engineering, Brno University of Technology, Technická 2, 61669 Brno, Czech Republic*

[3]*Department of Ceramics and Polymers, Brno University of Technology, Technická 2, 616 69 Brno, Czech Republic*

*E-mail: ja.arregi@ceitec.vutbr.cz, vojtech.uhlir@ceitec.vutbr.cz

**Contents:**

- Figures S1-S14
- Note S1: Wulff-Kaischev's theorem for a supported crystal

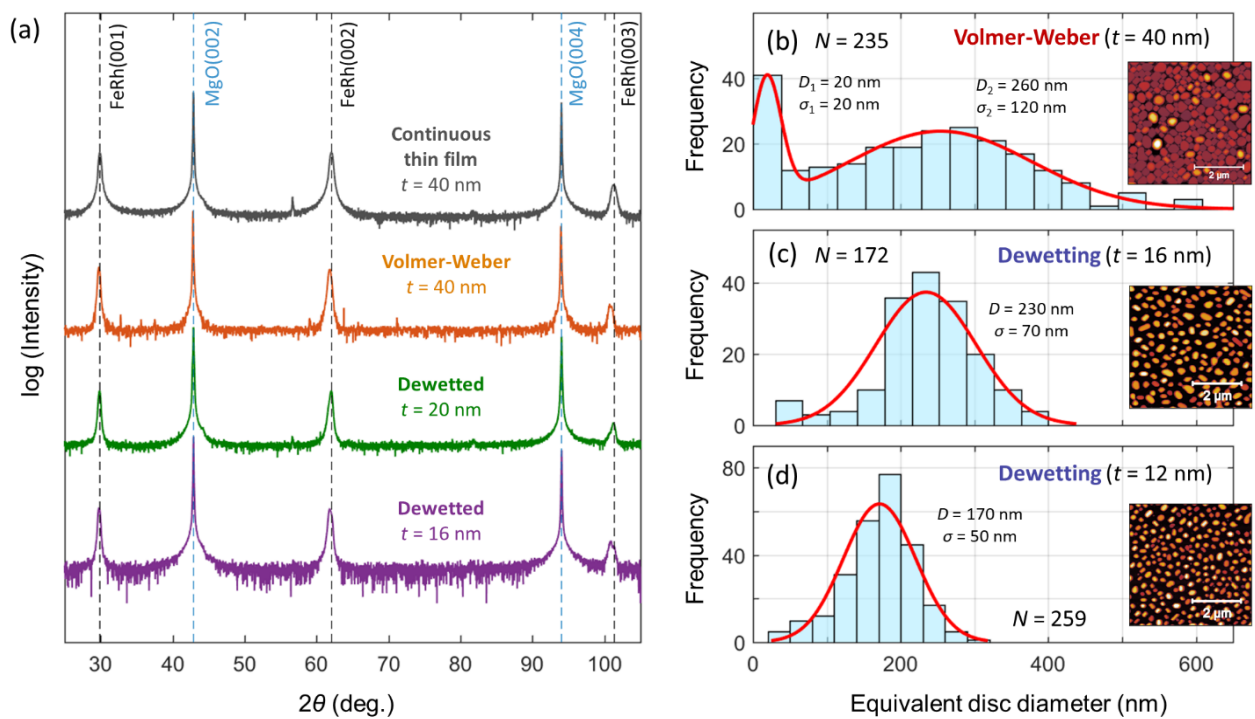

**Figure S1.** Crystallographic texture and size distribution analysis of self-assembled FeRh nanoislands. a) XRD $\theta$-$2\theta$ symmetric scans for FeRh nanoisland samples assembled via Volmer-Weber nucleation ($t$ = 40 nm) and solid-state dewetting ($t$ = 12, 16 nm) on MgO(001) substrates. All samples show a prominent (001) out-of-plane texture of the CsCl-type structure of FeRh. XRD data for a continuous FeRh film ($t$ = 40 nm) is shown for comparison. b)-d) show the size distribution of $N$ nanoislands evaluated from AFM topography data over a 5 × 5 µm² area (see insets). Histograms of the equivalent disc diameter are displayed for b) Volmer-Weber nucleated nanoislands with $t$ = 40 nm, as well as dewetted nanoislands with c) $t$ = 16 nm and d) 12 nm. Volmer-Weber nanoislands in b) show a bimodal distribution function featuring a large presence of small, sub-50 nm islands and a broad distribution of larger ones centered at the 260-nm-diameter size. Dewetted islands in c), d) exhibit narrower nanoisland size distributions with central diameter values 230 and 170 nm for $t$ = 16 and 12 nm, respectively. The red solid lines in b)-d) represent fits to normal distribution functions, with $D$ being the center diameter and $\sigma$ the standard deviation.

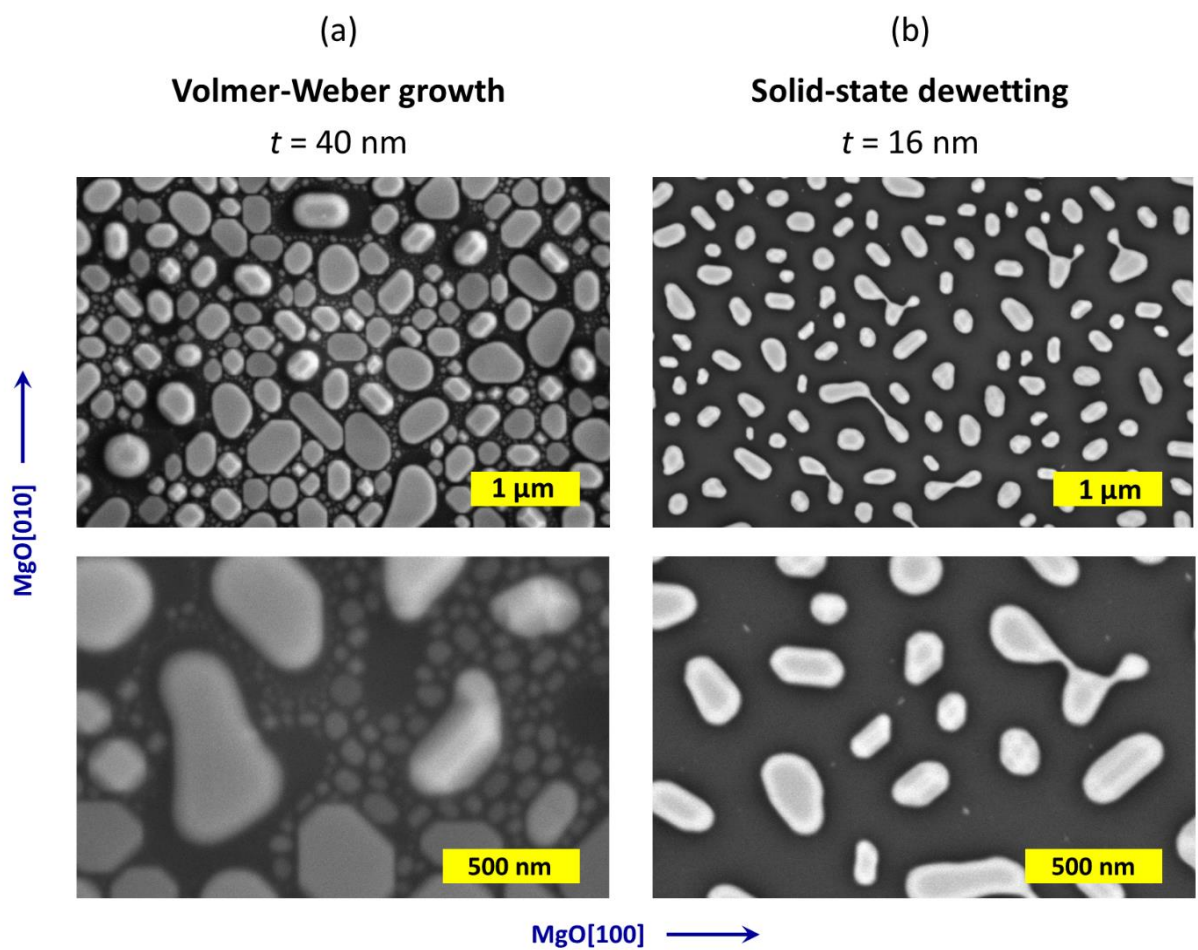

**Figure S2.** Electron microscopy images of FeRh nanoislands. Top-view SEM micrographs of FeRh nanoislands on MgO(001) formed via a) Volmer-Weber nucleation ($t = 40$ nm) and b) solid-state dewetting ($t = 16$ nm). The high magnification images in the bottom panel of a) reveal the ubiquitous presence of sub-50 nm nanoislands intercalated between larger ones (100-400 nm), pointing to a growth mechanism based on island nucleation. Nucleated FeRh islands show predominant faceting for the {100} crystallographic planes. The bottom panel in b) shows that dewetted nanoislands are well separated with no tiny islands in between. Dewetted islands also show a more diverse crystallographic faceting, giving them a more rounded appearance.

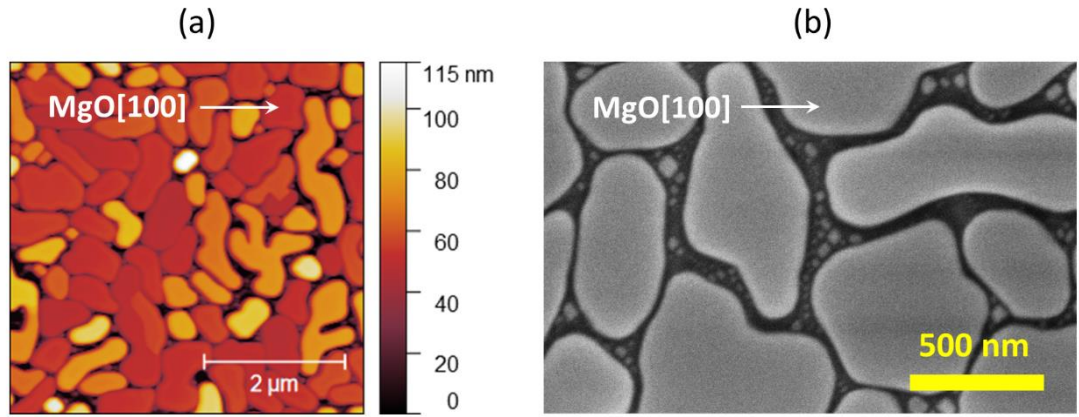

**Figure S3.** Morphology of high-temperature grown FeRh sample ($t = 40$ nm). a) AFM topography and b) SEM images of an FeRh sample grown at 1100 K and post-growth annealed at the same temperature for 80 min. The sample features islands with sizes in the ~ 0.5-1 µm range and more arbitrary shapes. The SEM image in b) reveals the existence of intercalated sub-50 nm nanoislands.

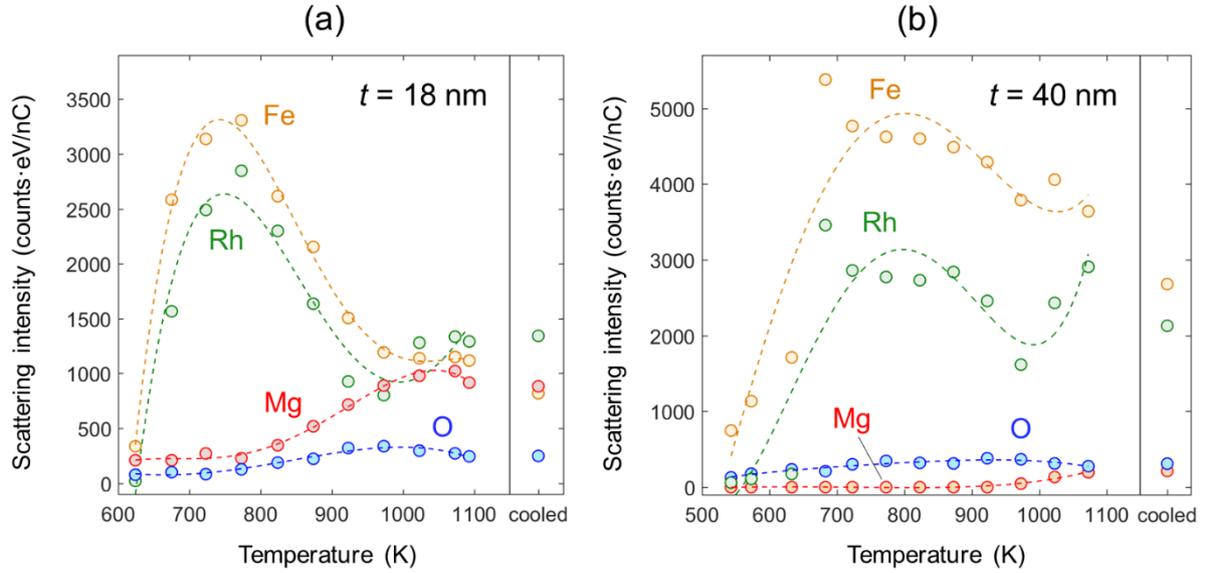

**Figure S4.** *In-situ* surface elemental analysis during dewetting of FeRh. Temperature-dependent LEIS intensity (integral of the energy-dependent peaks) for atomic Fe, Rh, Mg and O during *in-situ* annealing of FeRh films. Measurements are shown for a) 18-nm- and b) 40-nm-thick FeRh films on MgO(001). The dotted lines are a guide-to-the-eye obtained by a cubic spline of the measured data. The scattering intensity of each atomic species (representing their relative sample surface coverage) are shown while steadily ramping up the temperature at a rate of 9 K min$^{-1}$. The initial increase of the Fe and Rh atomic signals up to 700-800 K is indicative of surface dirt degassing (e.g., hydrocarbons). The subsequent (>800 K) decrease of the Fe and Rh scattering intensity is accompanied by an increase of the Mg and O signals, pointing towards dewetting of the FeRh thin film on the MgO substrate. These signatures are considerably strong for the 18-nm-thick film, while being less pronounced for the thicker 40 nm film, thus confirming the observed thickness dependence of the degree of dewetting in the FeRh/MgO system (Figure 1b of the main manuscript). The 'cooled' data points in the right-hand side of the graphs (performed upon cooling the sample down to 523 K right after reaching 1100 K) indicate that the dewetting-induced changes in the LEIS scattering intensity for Fe, Rh, Mg and O are permanent.

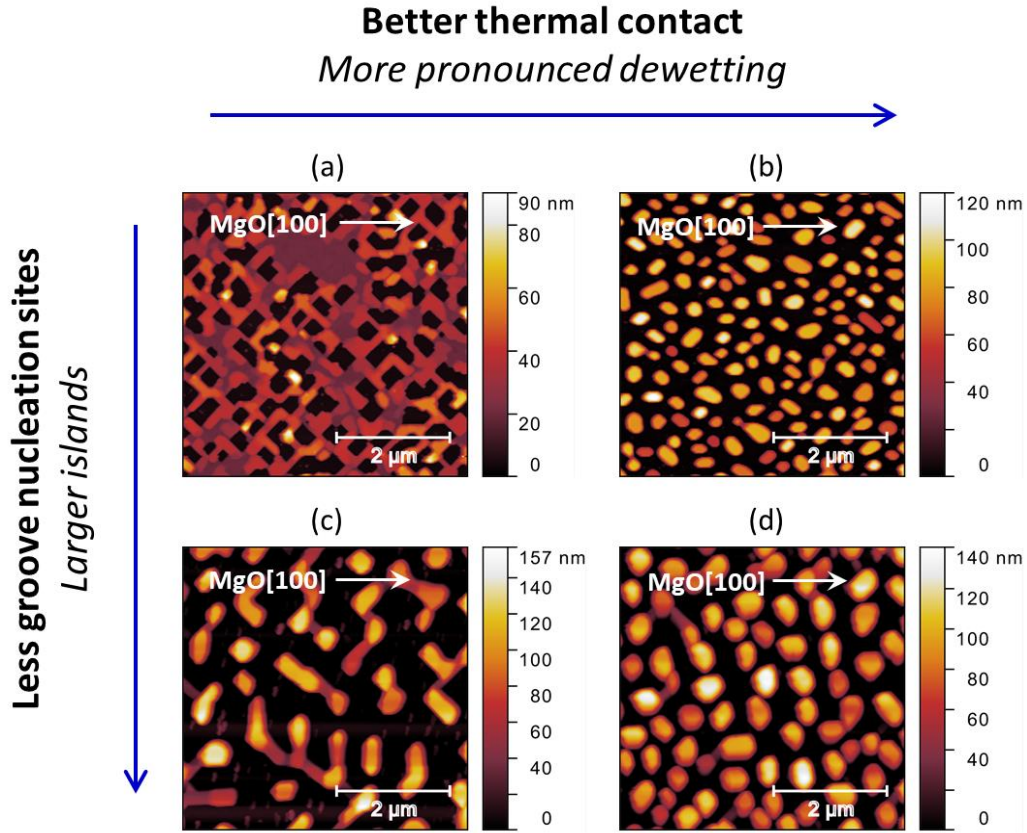

**Figure S5.** Impact of density and size of nucleated holes on the resulting size of nanoislands formed via solid state-dewetting. AFM topography scans over a $5 \times 5$ µm$^2$ area for four different dewetted FeRh(001)/MgO(001) samples with a nominal thickness of $t = 16$ nm. The samples were fabricated using the same nominal procedures for growth and annealing. We find that the final FeRh nanoisland morphology strongly depends on extrinsic factors such as the substrate-to-sample-holder thermal contact during fabrication (deposition and annealing), or the presence of defects and contamination on the substrate. We identify the general trends that impact the size and density of nanoislands. On the one hand, a better thermal contact of the substrate with the heater element during fabrication will lead to a more advanced dewetting stage and higher fragmentation of the film into nanoislands; compare a) and b), or c) and d). On the other hand, a larger number of defects and contaminants on the substrate lead to an increased presence of groove nucleation sites in the film, thus resulting in a higher density of void formation and smaller nanoislands; compare a) and c), or b) and d).

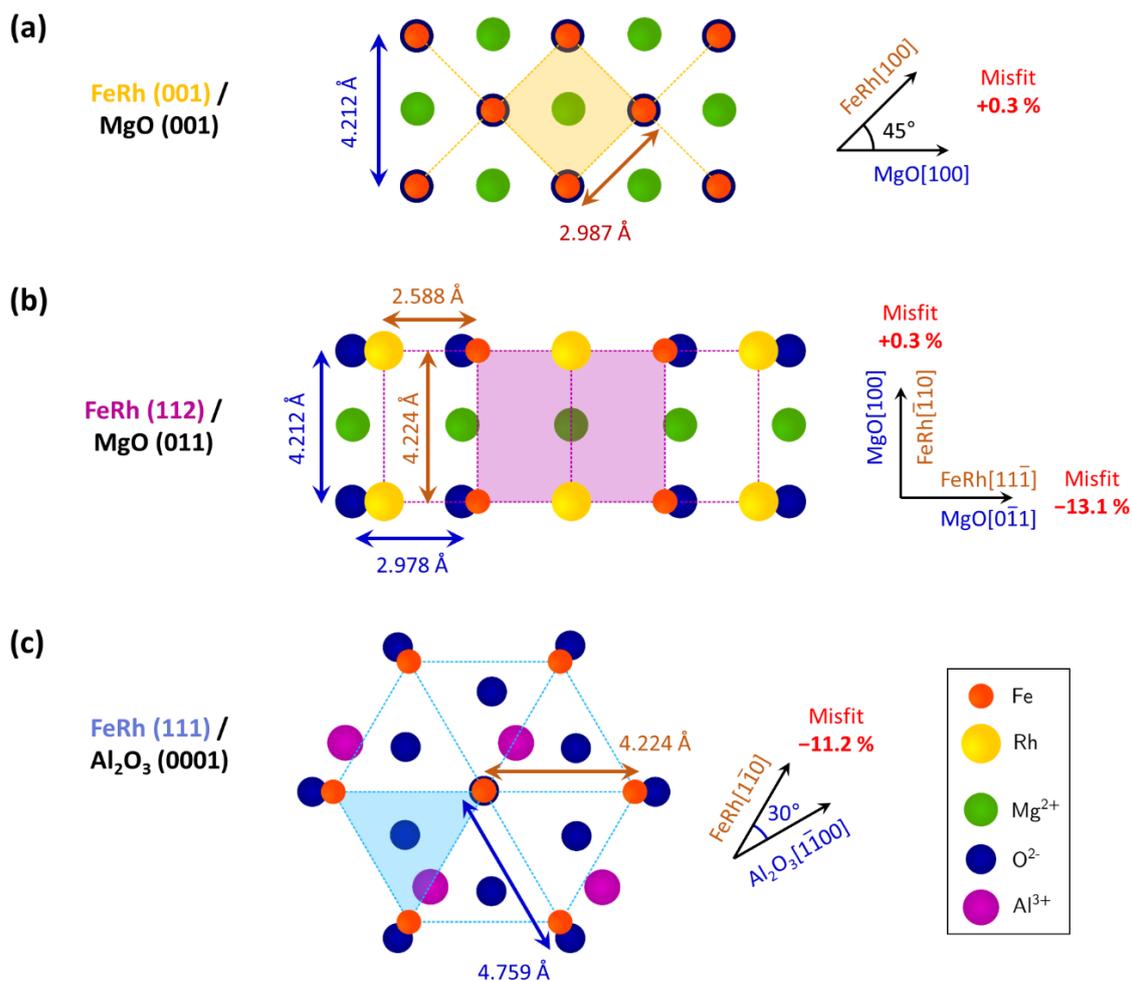

**Figure S6.** Epitaxial relations for FeRh growth on single crystal oxide substrates. Schematics of in-plane epitaxy for the a) FeRh(001)/MgO(001), b) FeRh(112)/MgO(011), and c) FeRh(111)/Al$_2$O$_3$(0001) systems. The characteristic lattice dimensions, orientations, and direction-dependent lattice misfit values are indicated in the graphics. The legend in the bottom-right panel indicates the colors used to represent the different atoms and ions in the lattice.

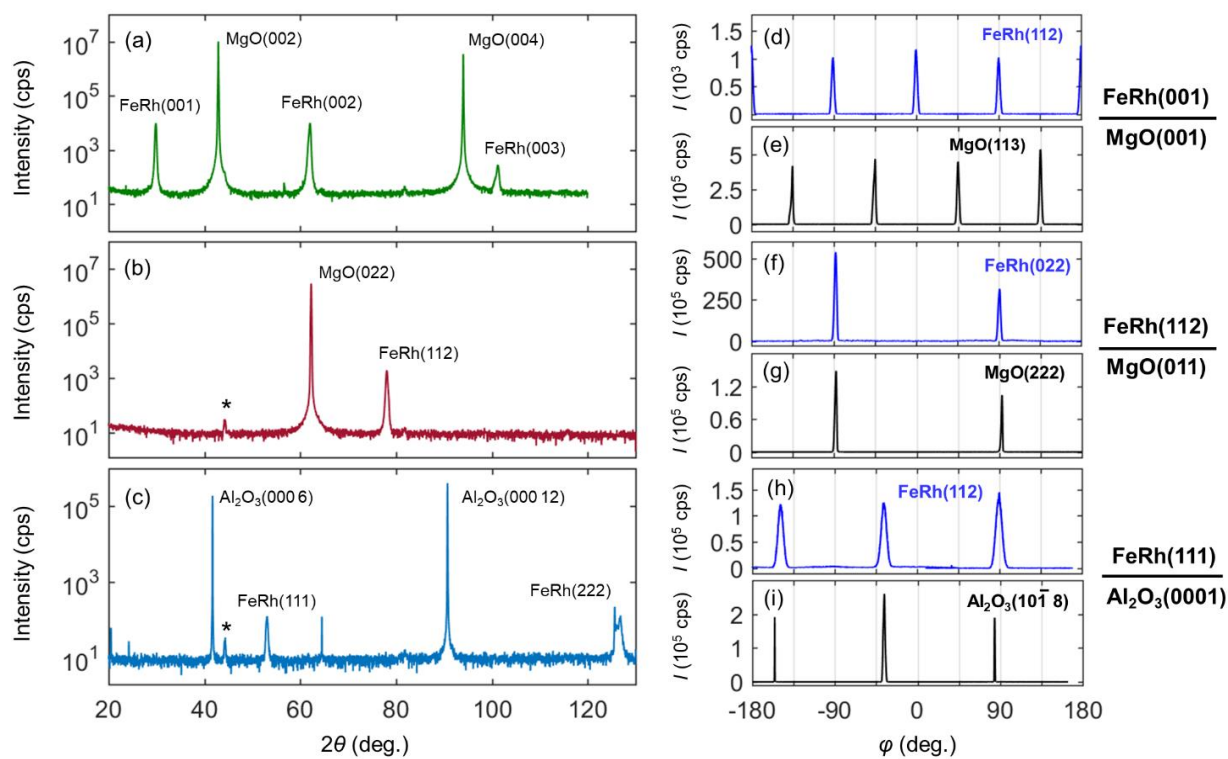

**Figure S7.** X-ray diffraction of self-assembled FeRh nanoislands ($t = 16$ nm) on different substrates formed via solid-state dewetting. The left panel exhibits symmetric $\theta/2\theta$ scans showing the out-of-plane crystallographic texture of the epitaxial a) FeRh(001)/MgO(001), b) FeRh(112)/MgO(011), and c) FeRh(111)/Al$_2$O$_3$(0001) nanoislands. The peaks labelled as '*' correspond to the sample holder stage. On the right panel, azimuthal $\varphi$-scans demonstrating the in-plane epitaxial relationships for the FeRh nanoislands on the three different substrate systems: d), e) FeRh(001)[100] || MgO(001)[110]; f), g) FeRh(112)[$\bar{1}$10] || MgO(001)[100]; h), i) FeRh(111)[11$\bar{2}$] || Al$_2$O$_3$(0001)[10$\bar{1}$0].

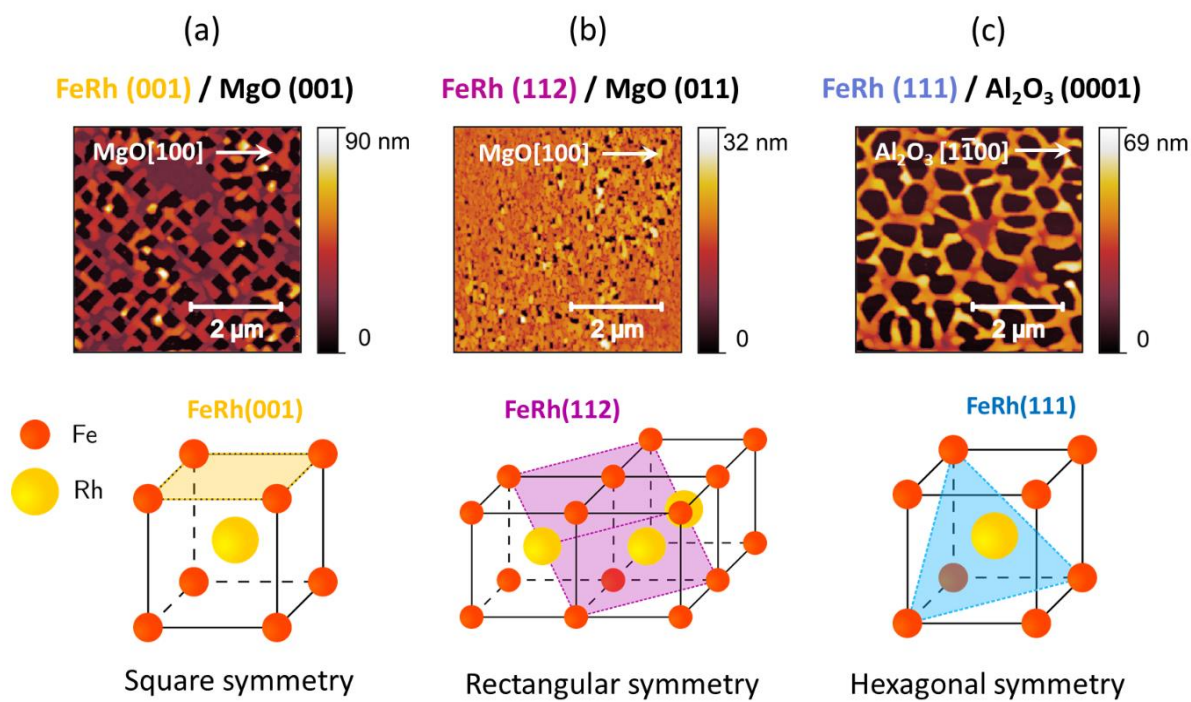

**Figure S8.** Morphology of partially dewetted epitaxial FeRh films. AFM microscopy images of FeRh films on a) MgO(001), b) MgO(011), and c) Al$_2$O$_3$(0001) substrates (with a nominal film thickness of 16, 16, and 12 nm, respectively) representing early stages of solid state dewetting. The samples were obtained upon shortening the annealing time or by the reduced thermal contact of the substrate with the heater stage (e.g., due to the insufficient clamping pressure when mounting the substrate). The morphology of the discontinuous films is characterized by the anisotropic nucleation of holes, these having a specific shape that is given by the crystallographic symmetry of the film.

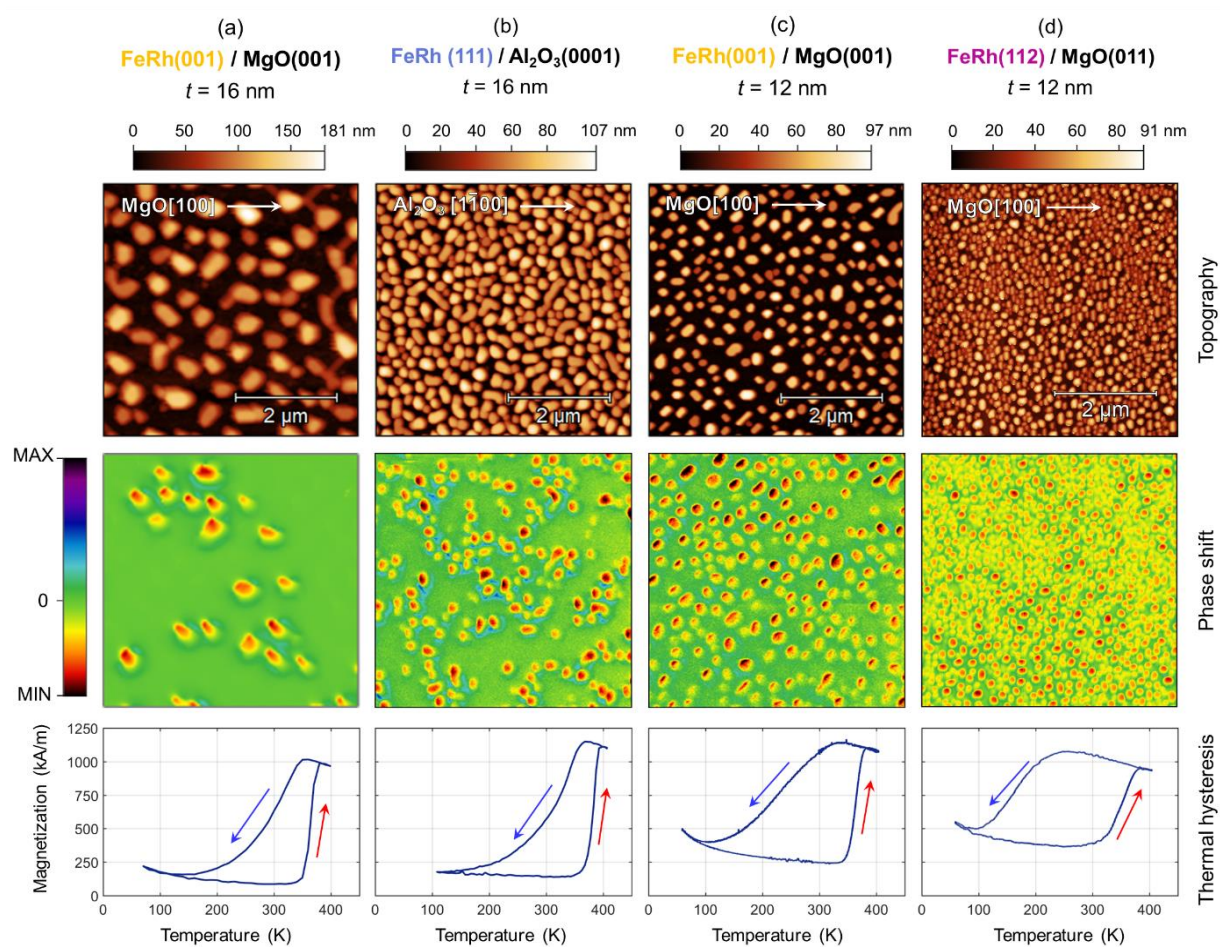

**Figure S9.** Morphology and magnetic properties of FeRh nanoislands assembled via solid-state dewetting on different substrates. a) FeRh(001)/MgO(001), $t$ = 16 nm; b) FeRh(111)/Al$_2$O$_3$(0001), $t$ = 16 nm; c) FeRh(001)/MgO(001), $t$ = 12 nm; d) FeRh(112)/MgO(011), $t$ = 12 nm. AFM topography images and room-temperature MFM measurements are shown over a 5 × 5 μm$^2$ area in the top and central rows, respectively. The bottom row of panels exhibit temperature-dependent magnetization measurements. Upon comparing FeRh(001) and FeRh(111) nanoislands with $t$ = 16 nm, it is seen that the FeRh(111) ones are of smaller characteristics size, as a result of the larger epitaxial mismatch of FeRh with the Al$_2$O$_3$(0001) substrate. Their phase transition characteristics are similar, with about a half or less of the nanoislands being FM at room temperature upon cooling from 400 K. In the case of the samples with $t$ = 12 nm in c), d), the thermal hysteresis is substantially broad during cooling, which is indicative of pronounced supercooling. The majority of islands are FM stabilized at room temperature upon cooling them down from 400 K. FeRh(112) nanoislands exhibit substantially smaller sizes (~100 nm) compared to the FeRh(001) ones, originating from the very large epitaxial mismatch of FeRh on MgO(011).

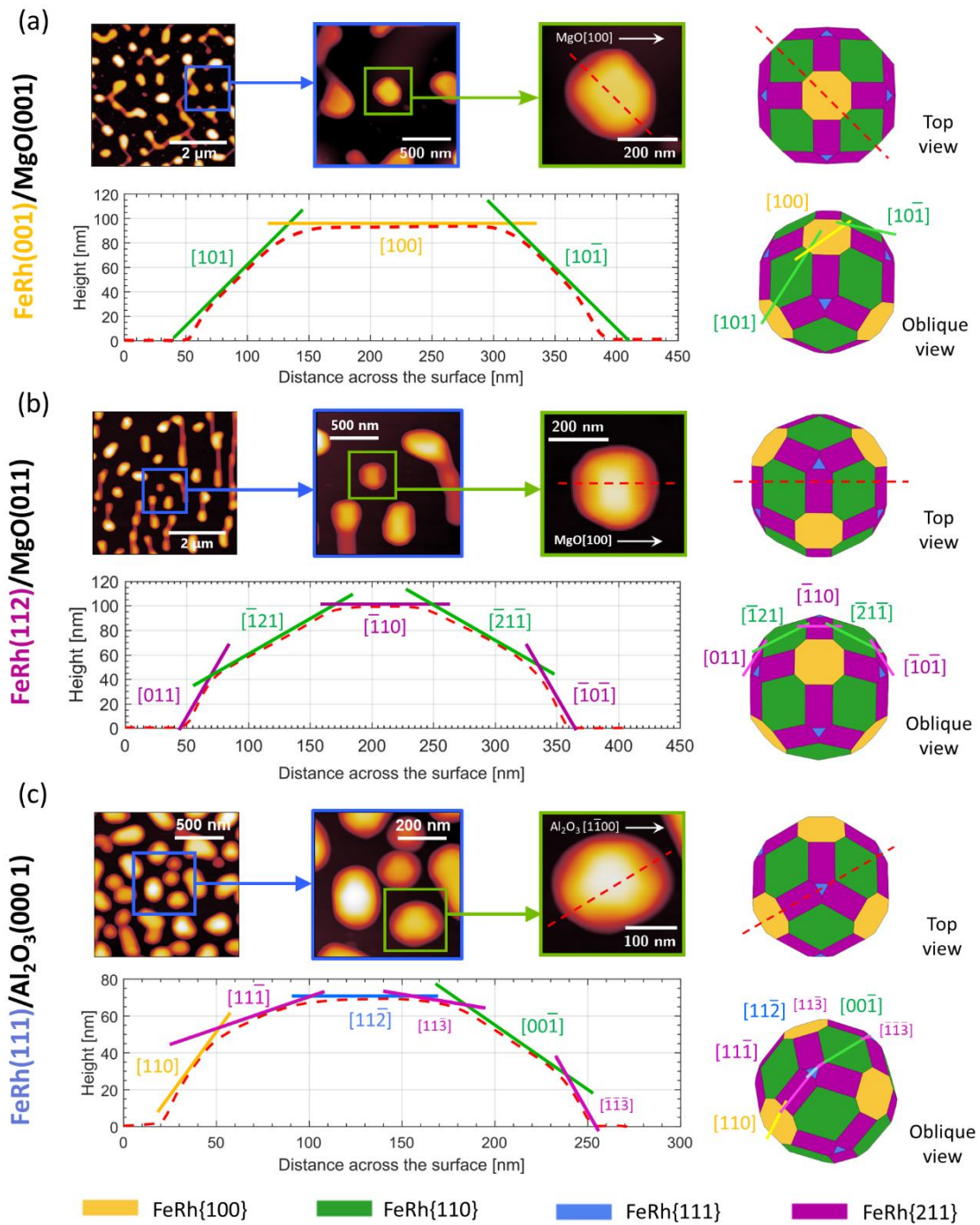

**Figure S10.** Crystal facet analysis of FeRh nanoislands via topographic line-scan profiles in the a) FeRh(001)/MgO(001), b) FeRh(112)/MgO(011), and c) FeRh(111)(Al$_2$O$_3$(0001) systems. AFM topography images (identical datasets as in Figure 2 of the main manuscript) are shown with accompanying line-scan profiles. The crystallographic directions along the line-scan are identified and indicated in the plots. The Wulff constructions including (100), (110), (111) and (211) planes are shown on the right hand side for each case, where the AFM line-scan trajectories are indicated in the top and oblique views. The colored legends at the bottom of the figure indicate the different crystal facet families.

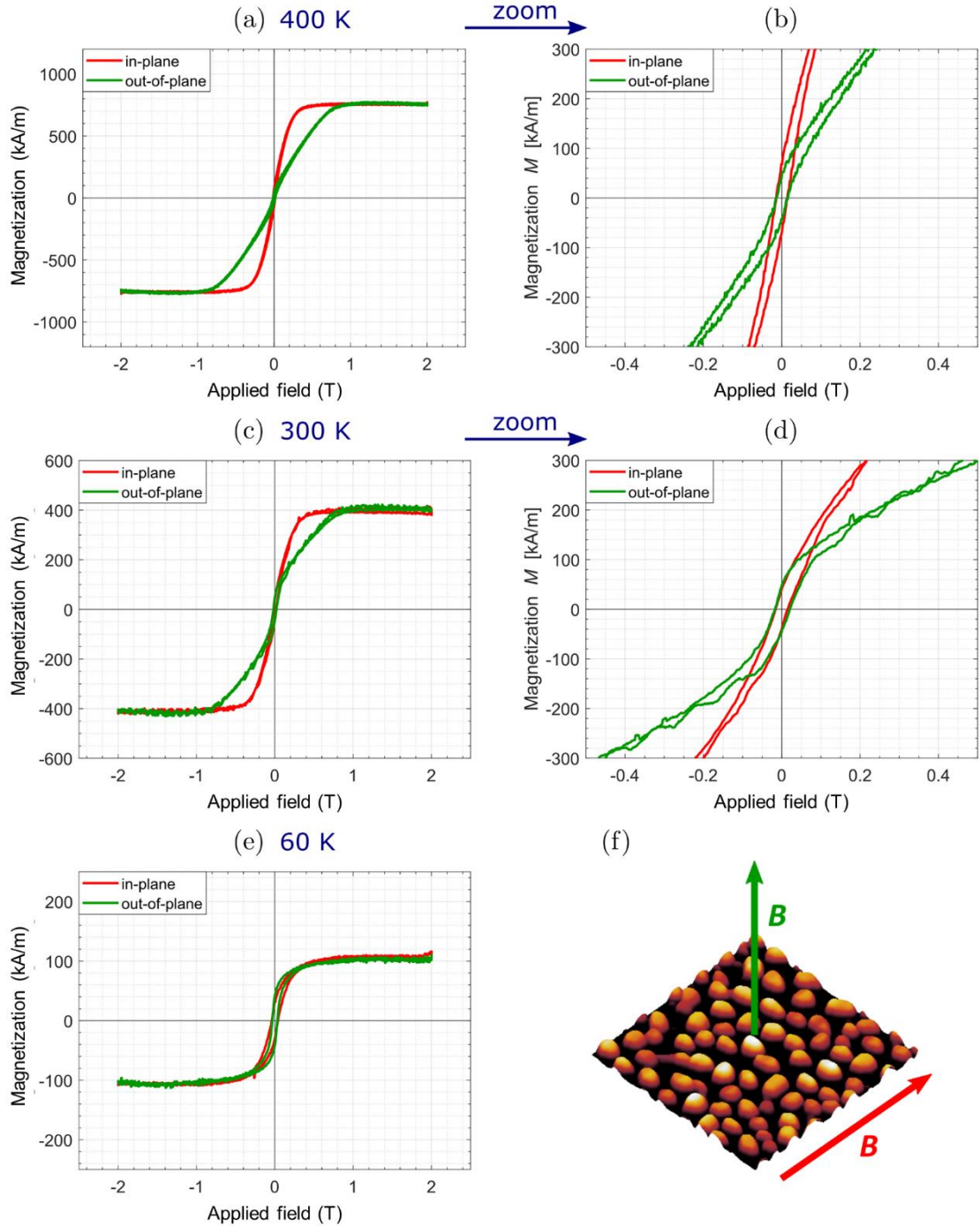

**Figure S11.** Field-dependent magnetization measured via VSM for FeRh(001) nanoislands with $t$ = 16 nm. Magnetization curves for in-plane and out-of-plane field configurations are shown a), b) at 400K; c), d) at 300K; e) at 60K. The plots in the right panel represent detailed views of magnetic loops displayed in the left panel. (f) 3D representation of the sample topography over a 5×5 μm$^2$ area, where the in-plane and out-of-plane magnetic field orientations are indicated by arrows.

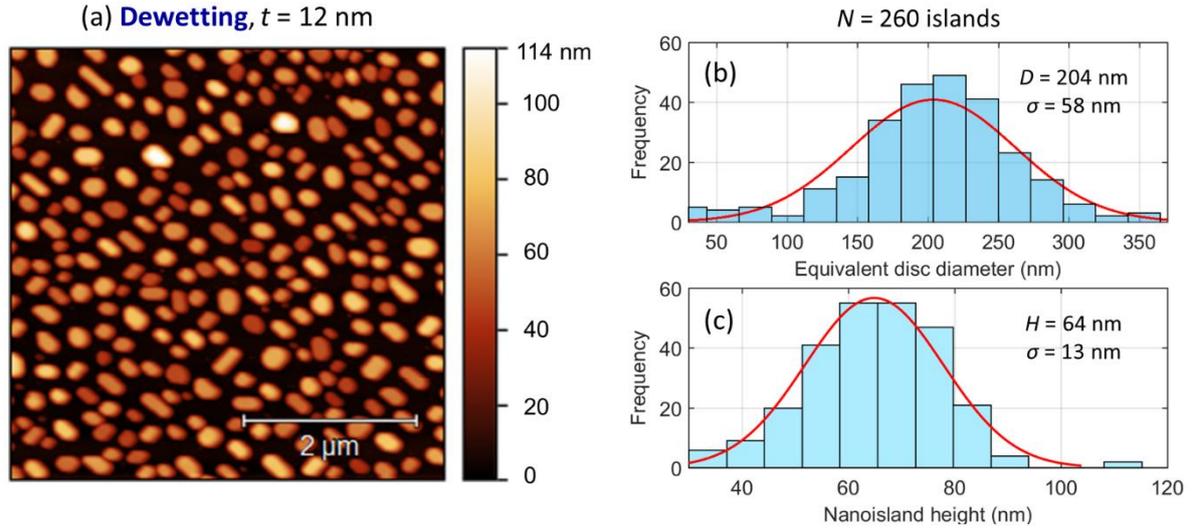

**Figure S12.** a) AFM image over a 5×5 μm² area of dewetted FeRh nanoislands ($t = 12$ nm) that were etched away from the MgO(001) substrate. b), c) Histograms of the equivalent disc diameter and height for the nanoislands ($N = 260$) shown in a). Central diameter and height values of $D = 204$ nm and $H = 64$ nm are obtained by fitting the histograms to a normal distribution function.

S-13

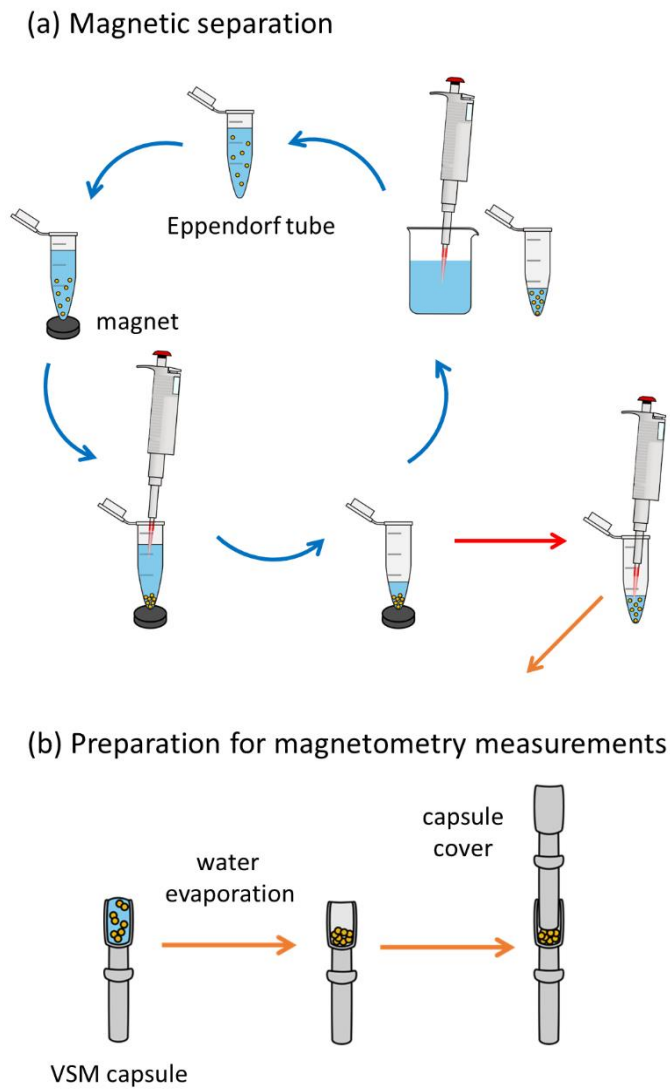

**Figure S13.** a) Schematics of nanoparticle separation process. The nanoparticles are initially dispersed or forming clusters in the EDTA disodium salt solution warmed to ~ 348 K within an Eppendorf tube. The solution is briefly heated to 363 K in order to induce the AF-to-FM transition in as many particles as possible. Afterwards, a large Nd-Fe-B permanent magnet providing a field of ~ 0.5 T at about ~ 3 mm from its surface is approached to the bottom part of the tube, causing the nanoparticles in the FM phase to agglomerate in this part. The excess solution from the top is removed using a pipette and the remaining content of the tube is subsequently diluted with deionized water. The process is repeated a few times in order to remove the EDTA disodium salt in the tube. b) Finally, the magnet is removed, and the reduced liquid volume is captured using a pipette, transferring it to a clean polypropylene capsule for VSM measurements. The water is let evaporate before closing the capsule prior to VSM measurements. The typical fraction of collected nanoparticles constituted about 40-to-70% of the supported nanoislands.

## Note S1: Wulff-Kaischev's theorem for a supported crystal

By assuming that the self-assembled FeRh nanoislands are elastically relaxed, the Wulff-Kaischev's theorem predicts the equilibrium shape of a supported crystal on a substrate, mathematically relating the occurrence and geometry of the crystal facets with their surface energies and the formation energy of the interface between the crystal and the substrate. The construction can be concluded considering that the following ratio remains constant for the different facets $i$ of the crystal

$$\frac{\gamma_i}{h_i} = \frac{\gamma_S - \gamma_{int}}{h_{int}}$$

(S1)

where $h_i$ is the distance from the nanocrystal's geometric center to a crystallographic facet $i$, $\gamma_i$ is the surface energy of the facet, $\gamma_S$ is the surface energy of the substrate, and $\gamma_{int}$ is the interface formation energy (see Figure S14). In addition, $h_{int}$ is the truncation height of the nanocrystal from its geometrical center. Here, the case $h_{int} > 0$ is considered, which leads to a supported nanocrystal truncated from above its center (see Figure S14).

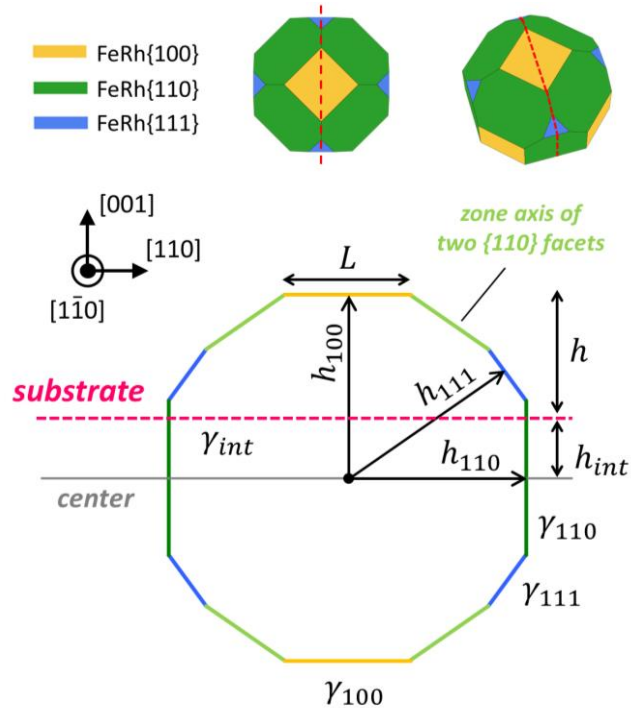

**Figure S14.** Depiction of the Winterbottom construction. On the top, schematics of a (001) oriented FeRh nanoisland truncated from above its center. The red dashed lines indicate the cutting plane ($[1\bar{1}0]$) along which the nanoisland morphological attributes were measured. On the bottom, definition of the nanoisland attributes used for the Wulff-Kaischev construction.

Considering {100}, {110} and {111} planes, Equation S1 can be rewritten as

$$\frac{\gamma_{100}}{h_{100}} = \frac{\gamma_{110}}{h_{110}} = \frac{\gamma_{111}}{h_{111}} = \cdots = \frac{\gamma_S - \gamma_{int}}{h_{int}}.$$

(S2)

For the FeRh nanoislands investigated here, $h_{int} > 0$ and thus $\gamma_{int} > \gamma_S \approx 1.17$ J m$^{-2}$. Setting $h_{int} = h_{100} - h$, with $h$ being the height of the nanoisland (see Figure S14), leads to

$$h = \frac{\gamma_{100} - \gamma_S + \gamma_{int}}{\gamma_{100}} h_{100}.$$

(S3)

For $\gamma_{int} < \gamma_S$, we have that $h < h_{100}$ (nanocrystal truncated from above).

Furthermore, we can also express the extent of the nanoisland cusp along the [110] direction (see Figure S14) as $L = 4h_{110} - 2\sqrt{2}h_{100}$, a relation that is independent of the nanoisland truncation height $\gamma_{int}$ and is obtained from evaluating the geometry of Wulff construction. Using Equation S2, $L$ can be rewritten as

$$L = \frac{4\gamma_{110} - 2\sqrt{2}\gamma_{100}}{\gamma_{100}} h_{100},$$

(S4)

and the combination of Equation S3 and S4 leads to

$$\frac{h}{L} = \frac{\gamma_{100} - \gamma_S + \gamma_{int}}{4\gamma_{110} - 2\sqrt{2}\gamma_{100}}$$

(S5)

which relates the measured $h/L$ of the nanoislands to the surface energies of the {100} and {110} facets of FeRh, the surface energy of the substrate $\gamma_S$, as well as the FeRh/MgO interface formation energy $\gamma_{int}$. Equation S5 can be rearranged to obtain

$$\gamma_{int} = (\gamma_S - \gamma_{100}) + \left(4\gamma_{110} - 2\sqrt{2}\gamma_{100}\right)\frac{h}{L}$$

(S6)

which is the same expression as in Equation 1 of the main manuscript. The $\gamma_{int}$ vs $h/L$ expression above can be interpreted as a straight line with a negative intercept ($\gamma_S < \gamma_{100}$) and a positive slope.